\documentclass{mn2e}
\usepackage{epsfig}
\usepackage{longtable}
\usepackage{mathrsfs}
\usepackage{graphicx}
\usepackage{amssymb}

\title[The protoplanetary disk Laques-Vidal 2 in Orion]{Chemical abundances in the protoplanetary disk LV\,2 (Orion): clues to the causes of the abundance anomaly in \hii\ regions}
\author[Y. G. Tsamis et al.]{Y. G. Tsamis$^{1, 2, 3}$\thanks{E-mail:
ytsamis@eso.org}, J. R. Walsh$^{2}$, J. M. V\'{i}lchez$^{1}$, and D.
P\'{e}quignot$^{4}$\thanks{Based on observations made with ESO telescopes at
the Paranal Observatory (078.C-0247(A); PI: Tsamis) and on
observations made with the NASA/ESA Hubble Space Telescope (PID 6034; PI: Walsh).}\\
$^{1}$Instituto de Astrof\'{i}sica de Andaluc\'{i}a (CSIC), Apartado 3004, 18080 Granada, Spain\\
$^{2}$European Southern Observatory, Karl-Schwarzschild-Str. 2, D-85748 Garching bei M$\ddot{u}$nchen, Germany\\
$^{3}$Department of Physics and Astronomy, The Open University, Walton Hall,
Milton Keynes MK7 6AA\\
$^{4}$LUTH, Observatoire de Paris, CNRS, Universit\'e Paris Diderot, 5 Place
Jules Janssen, 92190 Meudon, France}

\newcommand{\apj}{ApJ}
\newcommand{\apjs}{ApJS}
\newcommand{\aap}{A\&A}
\newcommand{\aj}{AJ}

\newcommand{\mnras}{MNRAS}
\newcommand{\hst}{{\it HST\/}}
\newcommand{\eld}{$N_{\rm e}$}

\newcommand{\crd}{$N_{\rm cr}$}
\newcommand{\elt}{$T_{\rm e}$}

\newcommand{\cmt}{cm$^{-3}$}

\newcommand{\cp}{C$^+$}
\newcommand{\cpp}{C$^{2+}$}

\newcommand{\op}{O$^+$}
\newcommand{\nep}{Ne$^+$}
\newcommand{\opp}{O$^{2+}$}

\newcommand{\np}{N$^+$}
\newcommand{\sulp}{S$^+$}
\newcommand{\sulpp}{S$^{2+}$}
\newcommand{\npp}{N$^{2+}$}

\newcommand{\nepp}{Ne$^{2+}$}

\newcommand{\clpp}{Cl$^{2+}$}
\newcommand{\arpp}{Ar$^{2+}$}
\newcommand{\arppp}{Ar$^{3+}$}
\newcommand{\fepp}{Fe$^{2+}$}
\newcommand{\foiii}{[O~{\sc iii}]}

\newcommand{\foi}{[O~{\sc i}]}
\newcommand{\foii}{[O~{\sc ii}]}
\newcommand{\fsii}{[S~{\sc ii}]}
\newcommand{\fsiii}{[S~{\sc iii}]}

\newcommand{\fnii}{[N~{\sc ii}]}

\newcommand{\fariv}{[Ar~{\sc iv}]}
\newcommand{\fcliii}{[Cl~{\sc iii}]}

\newcommand{\fneiii}{[Ne~{\sc iii}]}

\newcommand{\ffeiii}{[Fe~{\sc iii}]}

\newcommand{\oi}{O~{\sc i}}
\newcommand{\oii}{O~{\sc ii}}
\newcommand{\cii}{C~{\sc ii}}

\newcommand{\fciii}{C~{\sc iii}]}

\newcommand{\fariii}{[Ar~{\sc iii}]}

\newcommand{\hi}{H\,{\sc i}}
\newcommand{\hii}{H~{\sc ii}}
\newcommand{\hei}{He~{\sc i}}

\newcommand{\hp}{H$^+$}
\newcommand{\hep}{He$^+$}

\newcommand{\ha}{H$\alpha$}
\newcommand{\hb}{H$\beta$}
\newcommand{\hg}{H$\gamma$}
\newcommand{\hd}{H$\delta$}

\begin{document}

\date{Accepted XXX December XX. Received XXX December XX; in original form XXX October XX}

\pagerange{\pageref{firstpage}--\pageref{lastpage}} \pubyear{2002}

\maketitle

\label{firstpage}

\begin{abstract}

Optical integral field spectroscopy of the archetype protoplanetary disk LV\,2
in the Orion Nebula is presented, taken with the VLT FLAMES/Argus fibre array.
The detection of recombination lines of \cii\ and \oii\ from this class of
objects is reported, and the lines are utilized as abundance diagnostics. The
study is complemented with the analysis of \hst\ Faint Object Spectrograph
ultraviolet and optical spectra of the target contained within the Argus field
of view. By subtracting the local nebula background the intrinsic spectrum of
the proplyd is obtained and its elemental composition is derived for the first
time. The proplyd is found to be overabundant in carbon, oxygen and neon
compared to the Orion Nebula and the sun.

The simultaneous coverage over LV\,2 of the \fciii\ $\lambda$1908 and \foiii\
$\lambda$5007 collisionally excited lines (CELs) and \cii\ and \oii\
recombination lines (RLs) has enabled us to measure the abundances of \cpp\ and
\opp\ for LV\,2 with both sets of lines. The two methods yield consistent
results for the intrinsic proplyd spectrum, but not for the proplyd spectrum
contaminated by the generic nebula spectrum, thus providing one example where
the long-standing abundance anomaly plaguing metallicity studies of \hii\
regions has been resolved. These results would indicate that the standard
forbidden-line methods used in the derivation of light metal abundances in
\hii\ regions in our own and other galaxies underestimate the true gas
metallicity.

\end{abstract}


\section{Introduction}

LV\,2 (167-317; O'Dell \& Wen 1994) is found within the area defined by the
stars of the Trapezium cluster in the Orion Nebula (M42). Its first
characterization was made by Laques \& Vidal (1979) who imaged it as an
unresolved `nebular condensation' in the \ha, \hb\ and \foiii\ $\lambda$5007
emission lines and proposed that it is a partially ionized high-density
globule, a class apart from Bok globules or Herbig-Haro objects. Its true form
along with many similar objects, that of a \emph{protoplanetary disk} or
`proplyd', was revealed later in \hst\ images (O'Dell et al. 1993; O'Dell \&
Wen 1994). The currently accepted model for the proplyds is that they are
semi-ionized globules harbouring a pre-main sequence star girdled by an
accreting, and possibly planet-forming, disk (McCaughrean \& O'Dell 1996;
Johnstone et al. 1998; Bally et al. 2000). While molecular material
photoevaporated from the disk is shielded from ionizing Lyman continuum
radiation, proplyds that are situated within the main \hii\ region are being
exposed to both direct (stellar) and diffuse (scattered in the nebula) ionizing
photons with energies greater than 13.6\,eV and they become radiation-bounded,
surrounded by an ionization front.

Fig\,~1 shows an \hst\ image of LV\,2 in the light of \ha\ (O'Dell \& Wen
1994); its main features are the brightest portion of the ionization front
(`cusp') facing in the direction of $\theta^1$\,Ori\,C -- the main ionizing
source in M42 -- and an outflow trailing in the opposite direction. The proplyd
also possesses a bipolar jet whose brightest lobe is visible on Fig.\,~1
(Meaburn et al. 1993; Henney et al. 2002; Doi et al. 2004). A number of known
properties of LV\,2 are summarized by O'Dell et al. (2008). A previous integral
field spectroscopic study of LV\,2 was presented by Vasconcelos et al. (2005)
who observed the 5515 -- 7630\,\AA\ range with Gemini GMOS focusing on the
morphology and kinematics.


In this work we examine the physical properties of LV\,2 and those of its
immediate M42 vicinity via VLT optical integral field spectroscopy and \hst\
Faint Object Spectrograph UV and optical spectra placing emphasis on the
elemental composition of the proplyd. LV\,2 is in effect used as a probe of the
extinction, density, temperature, and chemical abundance structure of the Orion
Nebula at subarcsecond (milli-parsec) scales. Previous works have found no
evidence for substantial electron temperature (\elt) variations in M42 in the
plane of the sky (Rubin et al. 2003), but larger variations across the volume
of the nebula have been posited (O'Dell et al. 2003). The proplyds themselves
represent localized density enhancements reaching electron densities (\eld) of
$\sim$ 10$^6$ \cmt\ (Laques \& Vidal 1979; Henney et al. 2002) and most of them
are embedded in the diffuse \hii\ region (\eld\ $\sim$ 10$^3$--10$^4$ \cmt;
Esteban et al. 2004). Along lines of sight crossing proplyds the reliability of
classic temperature diagnostic ratios (\foiii\ $\lambda$4363/$\lambda$5007,
\fnii\ $\lambda$5754/$\lambda$6584) is compromised: spuriously high values of
\elt\ are derived when their inherently high densities are overlooked (Walsh \&
Rosa 1999; Rubin et al. 2003; O'Dell et al. 2003; Mesa-Delgado et al. 2008;
Tsamis et al. 2009).

Importantly, it is currently unknown what the abundances of elements in
proplyds are and whether they are different from those of the well-studied
background nebula or of the main sequence stars in Orion. The chemical
composition of planet-forming circumstellar envelopes is of great current
interest given the established positive correlation between host star
metallicity and the incidence of giant planetary companions (e.g. Gonzalez
1997; Santos, Israelian \& Mayor 2000; Gonzalez et al. 2001; Laws et al. 2003;
Neves et al. 2009).

Of equal importance is to examine what, if any, the role of proplyds might be
in the context of the long-standing abundance anomaly affecting studies of
galactic and extragalactic \hii\ regions, whereby larger abundances are derived
from the optical recombination lines (RLs) than from the collisionally excited
lines (CELs) of oxygen and carbon ions (e.g. Peimbert et al 1993; Tsamis et al.
2003a; Peimbert 2003; Esteban et al. 2004; Garc{\'{\i}}a-Rojas et al. 2004;
Tsamis \& P{\'e}quignot 2005; Ercolano 2009; Esteban et al 2009; Mesa-Delgado
et al 2008, 2010). Here the first detailed study of a proplyd geared towards
tackling these open issues is presented.

\setcounter{figure}{0}
\begin{figure}
\begin{center}
\includegraphics[scale=0.55, bbllx=5, bblly=15, bburx=700, bbury=400, angle=0]{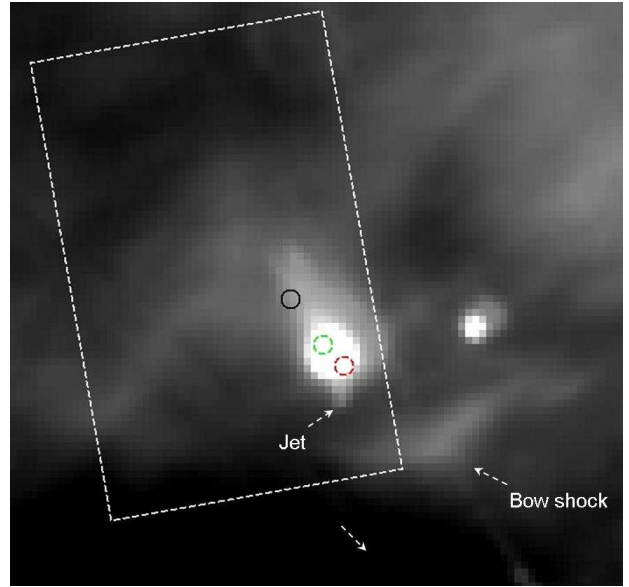}
\caption{The object near the centre of this \hst\ WFPC2 \ha\ $+$ \fnii\ image
(F656N filter) is the proplyd LV\,2 (167-317) at a scale of 0.0996$''$
pixel$^{-1}$. The overlaid rectangle is the 6.6$''$ $\times$ 4.2$''$ field of
view of the VLT Argus fibre array. The circles denote the 0.26$''$ diameter
{\it HST\/} FOS apertures corresponding to the `tip' (red), `core' (green), and
`tail' (black) positions discussed in the text. The arrow towards the bottom
right points in the direction of the ionizing source $\theta^1$\,Ori\,C. A
diffraction spike from $\theta^1$\,Ori\,C that was extending within the Argus
field of view was removed using interpolation from neighbouring pixels. The
proplyd outside the Argus field of view is 166-316. North is up and east is to
the left.}
\end{center}
\end{figure}

\setcounter{table}{0}
\begin{table}
\centering
\begin{minipage}{75mm}
\caption{Journal of observations.}
\begin{tabular}{@{}lccc@{}}
\hline
              Date                   &$\lambda$-range &Grating &Exp. time   \\
               (UT)                     &(\AA)           &        &(sec)               \\
\hline
&\multicolumn{3}{c}{VLT FLAMES$^a$}\\
2006/10/09                             &3062--4081      &LR1          &2$\times$150      \\
               "                      &3964--4567      &LR2          &3$\times$390      \\
               "                      &4501--5078      &LR3          &4$\times$206      \\
               2006/10/07               &5015--5831      &LR4          &3$\times$255    \\
               "                      &5741--6524      &LR5          &3$\times$255    \\
               "                      &6438--7184      &LR6          &3$\times$180    \\
&\multicolumn{3}{c}{{\it HST\/} FOS$^b$}\\
         1996/02/17               &1590--2312      &G190H        &200,200,100,70,30 \\
              "                 &2222--3277      &G270H        &90,58,80,40,60\\
              "                 &3235--4781      &G400H        &440,666,220,200,230\\
              "                 &4569--6818      &G570H        &880,1080,200,200,190\\
              "                 &6270--8500      &G780H        &100,120,80,60,100\\
\hline
\end{tabular}
\begin{description}
\item[$^a$] The Argus array was centered at (RA, Dec)$_{\rm JD2000}$ $=$
(05$^h$35$^m$16.857$^s$, $-$05$^{\circ}$23$'$15.03$''$) at a position angle of
$-$80 deg. \item[$^b$] The exposure times per grating sequentially refer to the
following coordinates (RA, Dec)$_{\rm JD2000}$: M42 filament
(05$^{h}$35$^{m}$16.119$^{s}$$\pm$0.06$''$,
$-$05$^{\circ}$24$'$09.82$''$$\pm$0.07$''$); M42 background
(05$^h$35$^m$16.220$^s$$\pm$0.06$''$,
$-$05$^{\circ}$24$'$09.47$''$$\pm$0.07$''$); LV\,2 tip
(05$^h$35$^m$16.737$^s$$\pm$0.135$''$,
$-$05$^{\circ}$23$'$16.48$''$$\pm$0.07$''$); LV\,2 core
(05$^h$35$^m$16.757$^s$$\pm$0.045$''$,
$-$05$^{\circ}$23$'$16.17$''$$\pm$0.07$''$); LV\,2 tail
(05$^h$35$^m$16.787$^s$$\pm$0.03$''$,
$-$05$^{\circ}$23$'$15.52$''$$\pm$0.07$''$).
\end{description}
\end{minipage}
\end{table}

\section{Observations and reductions}

Integral field spectroscopy of LV2 was performed on the 8.2-m VLT/UT2 Kueyen
during period 78 with the FLAMES Giraffe Argus array. A description of the
instrument can be found in Pasquini et al. (2002). A field of view of 6.6
$\times$ 4.2 arcsec$^{2}$ was used yielding 297 positional spectra in the 3062
-- 7184\,\AA\ range from six partially overlapping grating settings. The size
of the spatial resolution element was 0.31 $\times$ 0.31 arcsec$^{2}$,
corresponding to 123 $\times$ 123 AU$^{2}$ at the distance to M 42 (412 pc;
Reid et al 2009). The location of the Argus field on the Trapezium region of
M42 is shown in Fig.\,~1. The Argus data were reduced with the girBLDRS
pipeline developed by the Geneva Observatory which includes the cosmic-ray
removal, the flat-fielding, and the wavelength calibration via Th-Ar lamp
exposures (Blecha \& Simond 2004). The flux calibration was done within {\sc
iraf} using contemporaneous exposures of various spectrophotometric standards
for the grating settings [EG 21 (LR1), Feige 110 (LR2, LR3), LTT 7987
(LR4--6)]. Custom-made scripts were used to construct data cubes and spectral
line maps, and a $\chi^{2}$ minimization routine was used to fit Gaussians to
the emission lines (cf. Tsamis et al. 2008).

The observed data cubes are subject to atmospheric refraction as a function of
wavelength, known as differential atmospheric refraction (DAR). The direction
of DAR is along the parallactic angle at which the observation is made. The six
wavelength- and flux-calibrated data cubes for Giraffe low resolution modes LR1
to LR6 were corrected for this effect using the algorithm outlined by Walsh \&
Roy (1990); using the airmass information this procedure calculates fractional
pixel shifts for each monochromatic slice of a cube relative to a fiducial
wavelength (e.g. a strong emission line), shifts each slice with respect to the
orientation of the slit on the sky and the parallactic angle and recombines the
DAR-corrected data cube. The wavelength overlap of each LR setting does not
always allow a strong (e.g. \hi\ or \hei) line to be used to align each data
cube to correct for pointing differences between different grating set-ups. The
LV\,2 proplyd was the only compact feature that could be used to align the
images across the Argus field. The LR1, LR2, LR3 and LR6 cubes were aligned
using, respectively, the H$\epsilon$, H$\gamma$, H$\beta$ and H$\alpha$ lines;
the LR5 cube was aligned using \hei\ $\lambda$5876 by comparison with \hei\
$\lambda$4471 from the LR2 cube. For the LR4 range (5015--5830\,\AA), a sum of
the prominent (forbidden) lines -- [N~{\sc i}], \fcliii\ and \fnii\ -- was used
to compare with H$\beta$ to determine the shift. The shifts required to align
all the cubes were generally about 0.25 spaxels (0.08$''$) except for the LR1
cube whose shift was larger (0.20$''$). A result of this process is that the
corrected maps are no longer of identical extent, depending on the shifts
applied, but are suitable for line ratio maps to be constructed that have
fidelity on a spaxel-to-spaxel basis over their common imaged area.

LV\,2 and its surroundings were also observed with the \hst\ Faint Object
Spectrograph (FOS) in 1996 in programme 6034. The FOS Red detector (mode
FOS/RD) was employed with five gratings giving a total coverage from 1590 to
8500\,\AA\ with overlap between each spectral range (Keyes et al. 1995). The
0.3 arcsec single circular aperture was used which projects to 0.26 arcsec on
the sky in the aberrated \hst\ beam. Five positions on and in the near vicinity
of LV\,2 were observed. The positions were established by relative astrometry
on an \hst\ WFPC2 H$\alpha$ $+$ \fnii\ image taken in programme 5085 (PI: C. R.
O'Dell). Offsets were then specified from a star at $05^{h} 35^{m} 17^{s}.0,
-05^{\circ} 23' 34''.3$ of 15th mag. (in the $V$ band) selected from the
catalogue of Orion stars by Jones \& Walker (1988). This star was centred in
the FOS aperture with a series of acquisition peak-ups in successively smaller
apertures down to the observing aperture and then the offsets, measured from
this star to the observing positions on the WFPC2 image, were applied.

Three positions around LV\,2 were observed: the peak of LV2 (called `Core'),
0.6 arcsec southwest of the peak (called `Tip') and 1.0 arcsec northeast of the
peak (called `Tail'); see Fig.\,~1 for these aperture positions which are
contained within the Argus field of view. In addition two further positions, on
an \fsii-bright filamentary `ridge' of emission (`M42-filament') and a
background position 1.5 arcsec east of this filament (`M42-background'), were
observed; these fall outside the Argus field of view $\sim$55 arcsec southwest
from its centre. Table 1 lists the respective exposure times of all the grating
settings and positions. The standard pipeline reduction products were used for
the FOS data and emission line fluxes were determined by fitting Gaussians to
the extracted 1D flux and error spectra.

\section{Analysis of \hst\ FOS data}

\begin{figure*}
\centering \epsfig{file=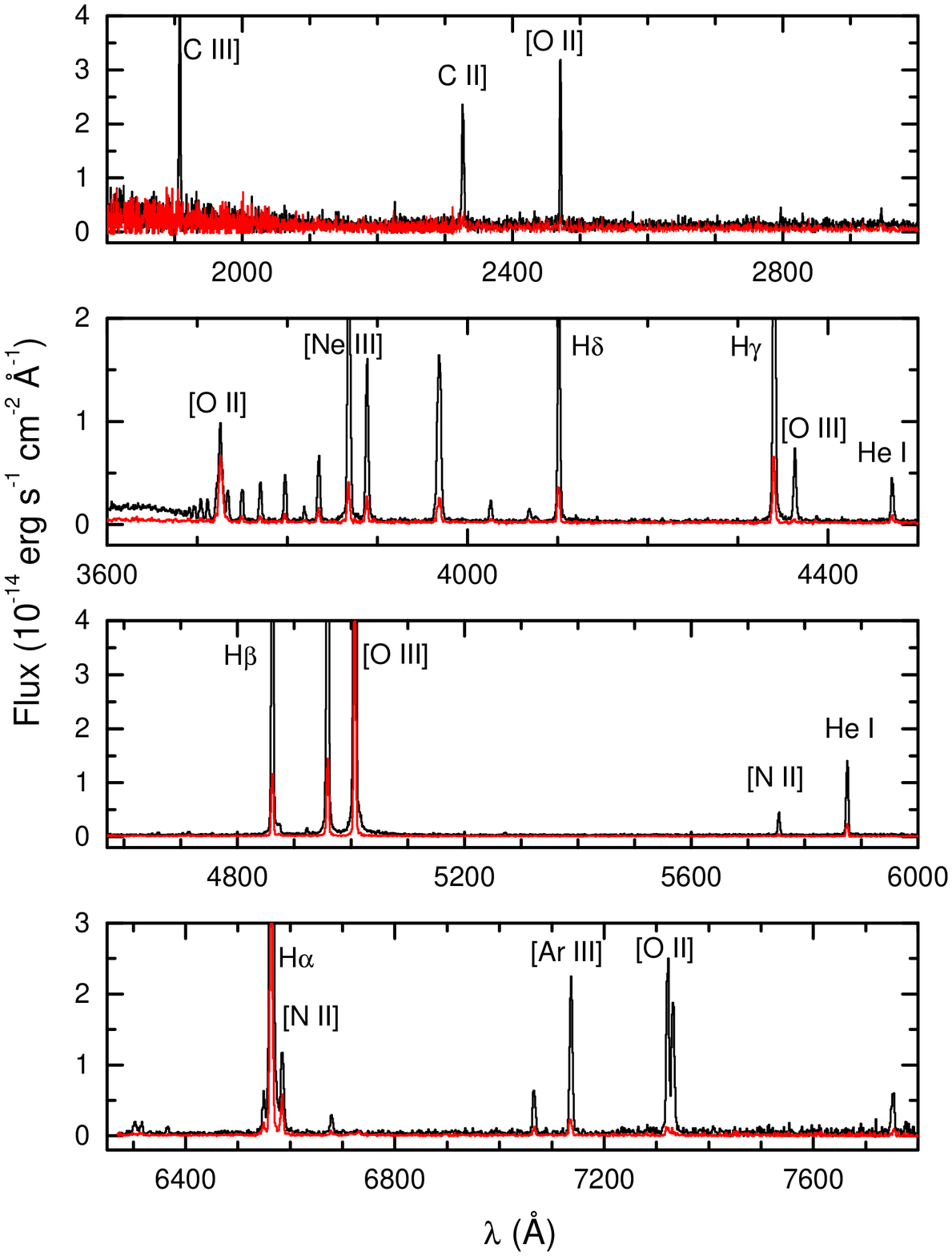, scale=.7, clip=, angle=0}
\caption{\hst\ FOS spectra of LV\,2 core (black) and tail (red) positions as
defined in Fig.\,~1 (not corrected for reddening).}
\end{figure*}

\setcounter{table}{1}
\begin{table*}
\caption{Line fluxes and physical conditions of M42 and LV\,2 positions from
{\it HST\/} FOS data.$^a$}
\begin{tabular}{lcccccc}
\noalign{\vskip3pt} \noalign{\hrule} \noalign{\vskip3pt}
                            &M42           &M42              &LV\,2      &LV\,2       &LV\,2        & LV\,2        \\
                            &filament    & background        &tip        & core     & tail      &   core $-$ tail\\
\noalign{\vskip3pt} \noalign{\hrule}\noalign{\vskip3pt} \noalign{\vskip3pt}

$c$(\hb)                    &0.77$\pm$0.07     & 0.71$\pm$0.07    &0.59$\pm$0.07    & 0.72$\pm$0.02     & 0.68$\pm$0.06      & 0.70$\pm$0.02 \\
$c$(\hb)                    &0.46$^{+0.16}_{-0.14}$ & 0.49$^{+0.45}_{-0.27}$     &1.01$^{+0.14}_{-0.12}$  &0.65$^{+0.09}_{-0.12}$   &0.78$^{+0.62}_{-0.31}$ &0.70$\pm$0.07 \\
$F$(\hb)                &0.385$\pm$0.02        & 0.340$\pm$0.02   &1.10$\pm$0.02    &3.53$\pm$0.04      &0.539$\pm$0.02  &2.99$\pm$0.04\\
\noalign{\vskip2pt}

$I$(\fciii\ $\lambda$1908)  & 16.0$\pm$5.5    & 13.1$\pm$8.7      & 42.4$\pm$7.2       & 75.2$\pm$5.8   & --      & 87.4$\pm$7.0  \\
$I$(C~\sc ii] $\lambda$2326) & 46.8$\pm$8.0    & 13.1$\pm$8.5      & 21.5$\pm$3.8       & 47.6$\pm$4.3   & $\lesssim$0.510  & 55.2$\pm$6.1 \\
$I$(\foii\ $\lambda$2470)   & 24.8$\pm$4.4    & 12.4$\pm$4.8      & 15.7$\pm$2.3       & 34.9$\pm$2.5   & 7.10$\pm$3.80    & 39.4$\pm$3.8 \\
$I$(\foii\ $\lambda$3726$+$29)   & 171$\pm$5     & 143$\pm$6       & 34.2$\pm$1.6       & 17.2$\pm$0.4   & 75.9$\pm$3.1    & 8.92$\pm$0.86 \\
$I$(\fneiii\ $\lambda$3869) &12.7$\pm$1.3   &13.7$\pm$2.0       &43.8$\pm$1.6           &45.3$\pm$0.9   &30.6$\pm$1.8 &47.6$\pm$2.1   \\
$I$(\hei\ $\lambda$4471)    &3.70$\pm$0.8    &3.65$\pm$1.13        &4.69$\pm$0.59           &4.47$\pm$0.30    &4.34$\pm$0.89     &4.49$\pm$0.39   \\
$I$(\foiii\ $\lambda$4959)  &68.4$\pm$3.6  &99.5$\pm$4.8      & 144$\pm$3        & 133$\pm$1    & 121$\pm$3     & 135$\pm$2   \\ 
$I$(\hei\ $\lambda$5876)    &10.7$\pm$1.7    & 12.1$\pm$2.1      &13.7$\pm$0.95           &14.0$\pm$0.5    &13.5$\pm$1.2      & 14.1$\pm$0.6       \\
$I$(\fsiii\ $\lambda$6312)   &1.07$\pm$0.04   & --        & 1.47$\pm$0.40          &1.96$\pm$0.25       &3.18$\pm$1.69    &2.00$\pm$0.35  \\
$I$(\fnii\ $\lambda$6584)   & 104$\pm$7   & 60.0$\pm$6.8       & 18.3$\pm$1.4       & 18.3$\pm$0.8   & 41.6$\pm$3.8    & 14.5$\pm$1.0   \\
$I$(\hei\ $\lambda$6678)    &3.22$\pm$1.6    &2.18$\pm$1.04        &3.29$\pm$0.67           &4.06$\pm$0.39    &3.64$\pm$1.48     & 4.13$\pm$0.45  \\
$I$(\fsii\ $\lambda$6731)   & 12.3$\pm$2.6   & 2.06$\pm$0.08       & 1.55$\pm$0.44       & 0.512$\pm$0.193  & 4.13$\pm$1.52    & -- \\
$I$(\fariii\ $\lambda$7135)  &8.73$\pm$2.9   &14.5$\pm$3.0       &41.4$\pm$1.9           &24.1$\pm$1.5  &15.6$\pm$2.2         &25.5$\pm$1.9 \\
$I$(\foii\ $\lambda$7320$+$30)  & 25.6$\pm$5.4    & 13.5$\pm$4.1      & 40.8$\pm$5.9       & 46.6$\pm$2.5   & 11.2$\pm$0.7    & 52.3$\pm$5.5 \\
\foiii\ $\lambda$4959/$\lambda$4363  &63.9$\pm$6.3 & --   &27.6 $\pm$3.0  & 17.4$\pm$0.9       &63.5$\pm$21.4   & 15.7$\pm$0.9  \\
\fnii\ $\lambda$6584/$\lambda$5755    & 31.3$\pm$8.6  & --   & 12.7$\pm$1.7   & 4.22$\pm$0.21   & 28.2$\pm$11.2     & 3.03$\pm$0.32       \\
\fsii\ $\lambda$6731/$\lambda$6716  & 2.53$\pm$1.06 & 1.75$\pm$0.26   &3.73$\pm$1.68       &--   &1.78$\pm$1.54  & --\\

\noalign{\vskip2pt}
\eld\ \fsii\ (\cmt)    &$>$2300   & 4300$^{+5500}_{-2100}$       &$\gtrsim$10$^{4}$      &--  &4800:     &  --     \\
\eld\ \foii\ (\cmt)$^b$    &1.7$^{+0.4}_{-0.2}$$\times$10$^4$   & 1.1$^{+0.8}_{-0.4}$$\times$10$^4$       &4.5$^{+0.7}_{-0.4}$$\times$10$^4$      &2.0$^{+4.2}_{-0.3}$$\times$10$^{5}$  &1.0$^{+0.9}_{-0.3}$$\times$10$^4$    & 6.6$^{+0.1}_{-0.1}$$\times$10$^{5}$   \\
\eld\ \fnii\        (\cmt) &6.5$^{+2.2}_{-1.9}$$\times$10$^4$ & --        &1.4$^{+0.14}_{-0.9}$$\times$10$^{5}$    &5.7$^{+0.2}_{-3.8}$$\times$10$^{5}$    & 7.9$^{+7.1}_{-6.2}$$\times$10$^4$  & 1.0$^{+0.1}_{-0.3}$$\times$10$^{6}$      \\
\elt\ \foiii\ (K)          & 8850$\pm$250      & --        &10950$\pm$550        &10700$\pm$1300      & 8950$^{+1550}_{-1250}$   &9050$\pm$650 \\


\noalign{\vskip3pt} \noalign{\hrule}\noalign{\vskip3pt}
\end{tabular}
\begin{description}
\item[$^a$] $c$(\hb) in row 1 were deduced from the \hi\ Balmer ratios, in row
2 from the \foii\ $\lambda$2470/$\lambda$7320+30 ratio. The former were
adopted; $F$(\hb)'s correspond to observed fluxes within a 0.26$''$ diameter
aperture (in units of $\times$10$^{-13}$ erg\,cm$^{-2}$\,s$^{-1}$);
$I$($\lambda$) denotes dereddened intensities in units of \hb\ $=$ 100;
\item[$^b$] Derived from the $I$($\lambda$3726)/$I$($\lambda$2470) ratio for
\elt\ $=$ \elt(\foiii); for `M42 background' 8400\,K was adopted from the Argus
background spectrum.



\end{description}
\end{table*}

The FOS spectra were used for the computation of the physical conditions and
abundances of the targeted positions, and as a means to independently check the
reliability of the FLAMES spectrophotometry for the overlapping LV\,2
positions. Spectra of the LV\,2 core and tail observations are shown in
Fig.\,~2. In Table 2 we present the dereddened fluxes for the strongest lines
detected at each position, along with the electron densities and temperatures
derived from standard diagnostic line ratios.\footnote{These were computed
using the code \textsf{EQUIB} developed at University College London.} The
adopted logarithmic reddening coefficient, $c$(\hb), was obtained from the
\ha/\hb, \hg/\hb\ and \hd/\hb\ flux ratios using the modified Cardelli, Clayton
\& Mathis (1989; CCM) law from Blagrave et al. (2007) with a total to selective
extinction ratio $R_{\rm V}$ $=$ 5.5 applicable to M42. Values of $c$(\hb) were
also deduced from a comparison of the observed \foii\
$\lambda$2470/$\lambda$7320+30 line ratio to the almost invariant theoretical
value of 0.75 taken from Zeippen (1982) (Table 2). These measurements typically
suffer from larger uncertainties than the Balmer decrement. It is also possible
that stellar light scattered within the nebula selectively enhances the UV/blue
lines (e.g. O'Dell \& Harris 2010) leading to lower reddening deduced from the
UV/red \foii\ ratio for the M42 filament and background positions than from the
Balmer decrement; in view of this the $c$(\hb) values from \foii\ were not used
in this analysis.

Electron densities were derived from the \fsii\ $\lambda$6731/$\lambda$6716
doublet ratio and the \foii\ $\lambda$3726/$\lambda$2470 ratio. The former
diagnostic is not sensitive to densities $\gtrsim$10$^4$ \cmt, while the latter
is sensitive up to a few times 10$^6$ \cmt\ (e.g. Osterbrock 1989). At the same
time the \foii\ ratio is somewhat \elt-sensitive: typically, higher \foii\
densities are returned for lower temperatures. The temperature applicable to
the \opp\ zone was derived from the \foiii\ $\lambda$5007/$\lambda$4363 ratio.
For each FOS position the mean \elt(\foiii) is quoted in Table 2 with a range
corresponding to an adopted upper and lower density boundary. Those boundaries
were taken from the positions where the \foii\ and \fnii\ curves cross the
\foiii\ curve on the (\elt, \eld) diagnostic plane for the M42 filament, and
LV\,2 tip and tail positions. The \fnii\ $\lambda$5754/$\lambda$6584 ratio,
just like the \foii\ $\lambda$3726/$\lambda$2470 ratio, is mostly density
sensitive for these positions. For LV\,2 core the upper density boundary was
taken to be 10$^6$ \cmt\ (see Section~4.2). The auroral \fnii\ $\lambda$5754
and \foiii\ $\lambda$4363 lines are not detected in the M42 background spectrum
and hence no \elt\ is available for it.

\section{Analysis of VLT Argus data}

\begin{figure*}
\centering \epsfig{file=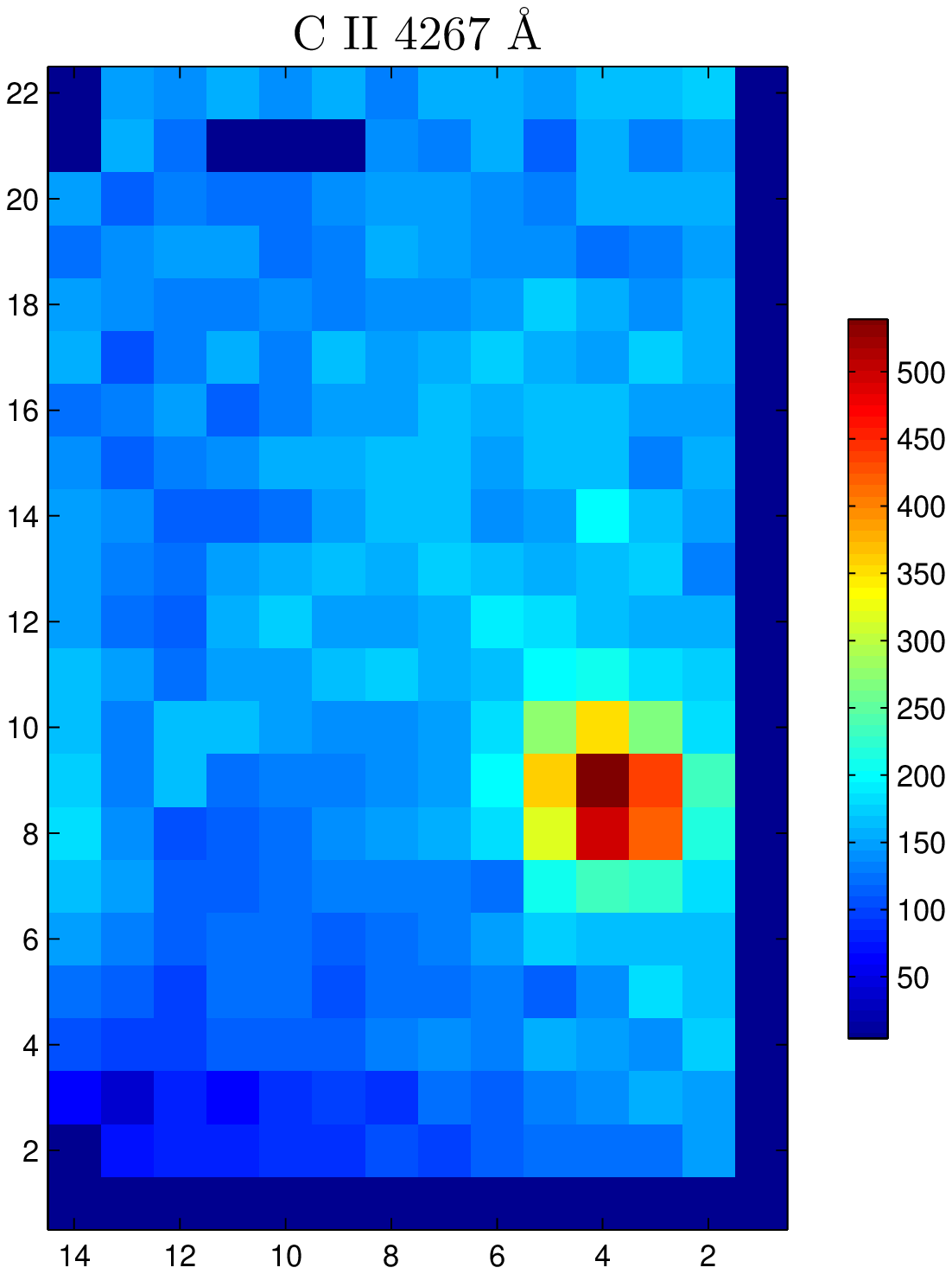, scale=0.42, clip=, bbllx=170, bblly=175,
bburx=500, bbury=675} \epsfig{file=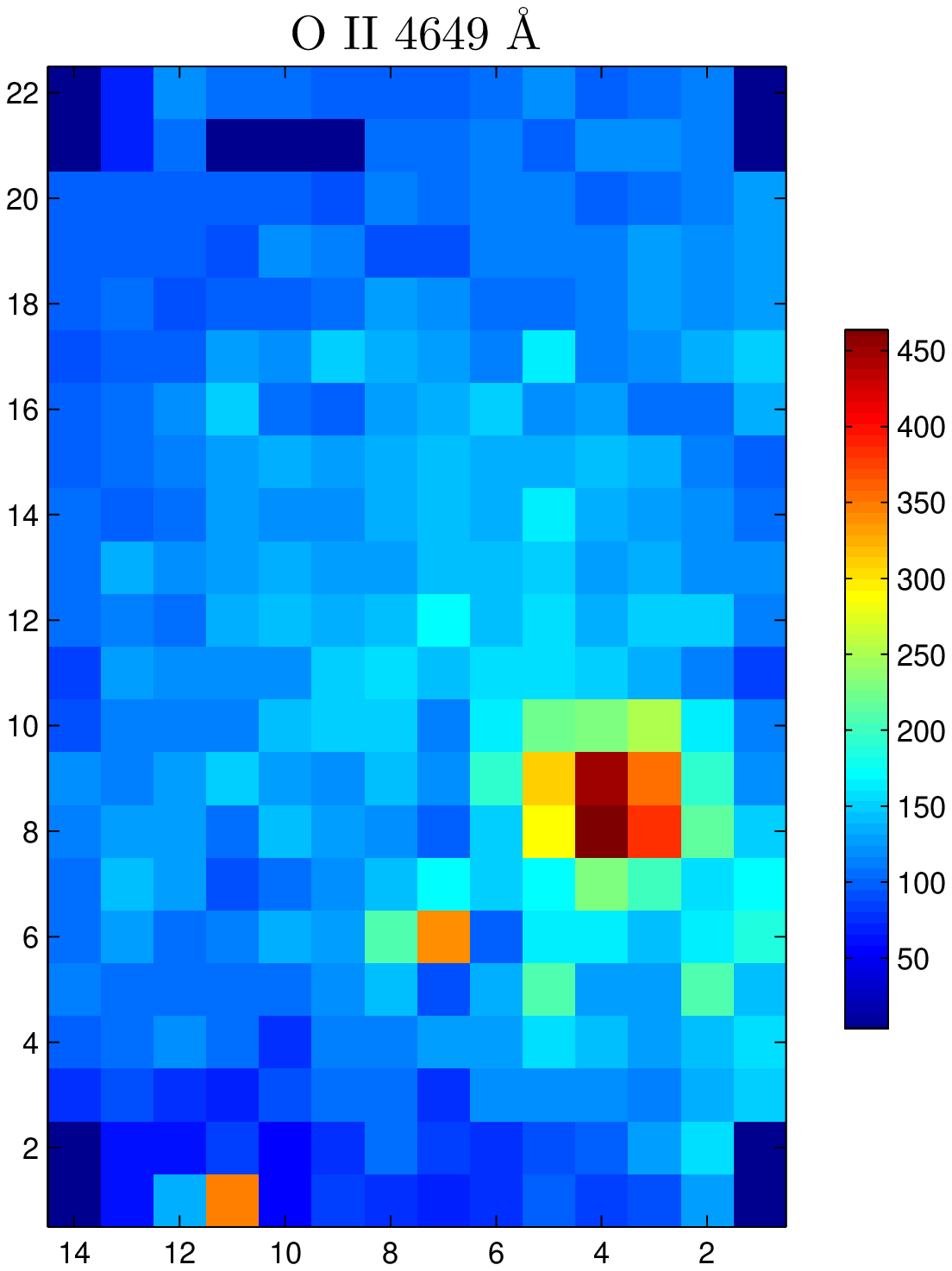, scale=0.42, clip=, bbllx=170,
bblly=175, bburx=500, bbury=675} \epsfig{file=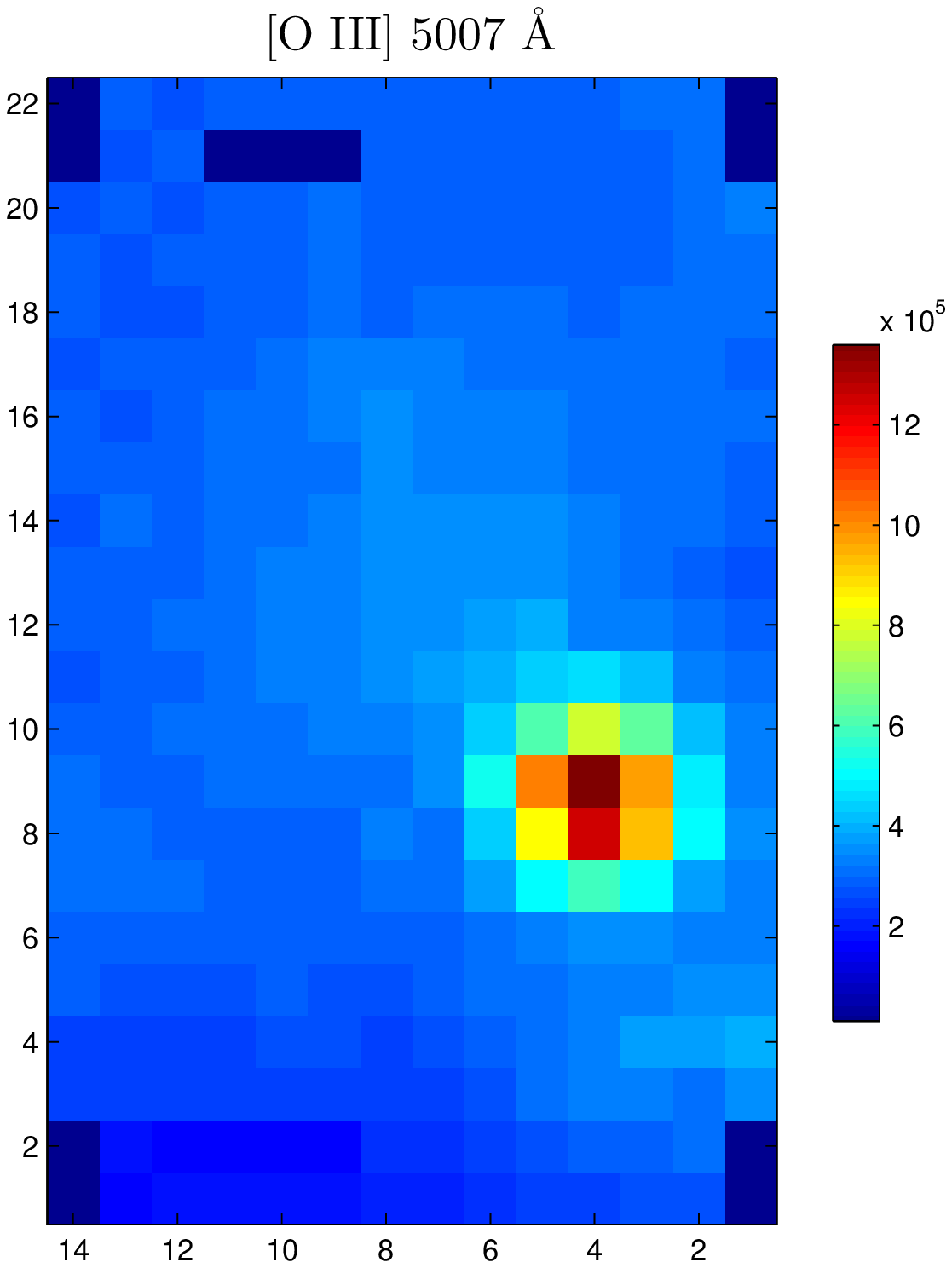, scale=0.42, clip=,
bbllx=170, bblly=175, bburx=500, bbury=675} \epsfig{file=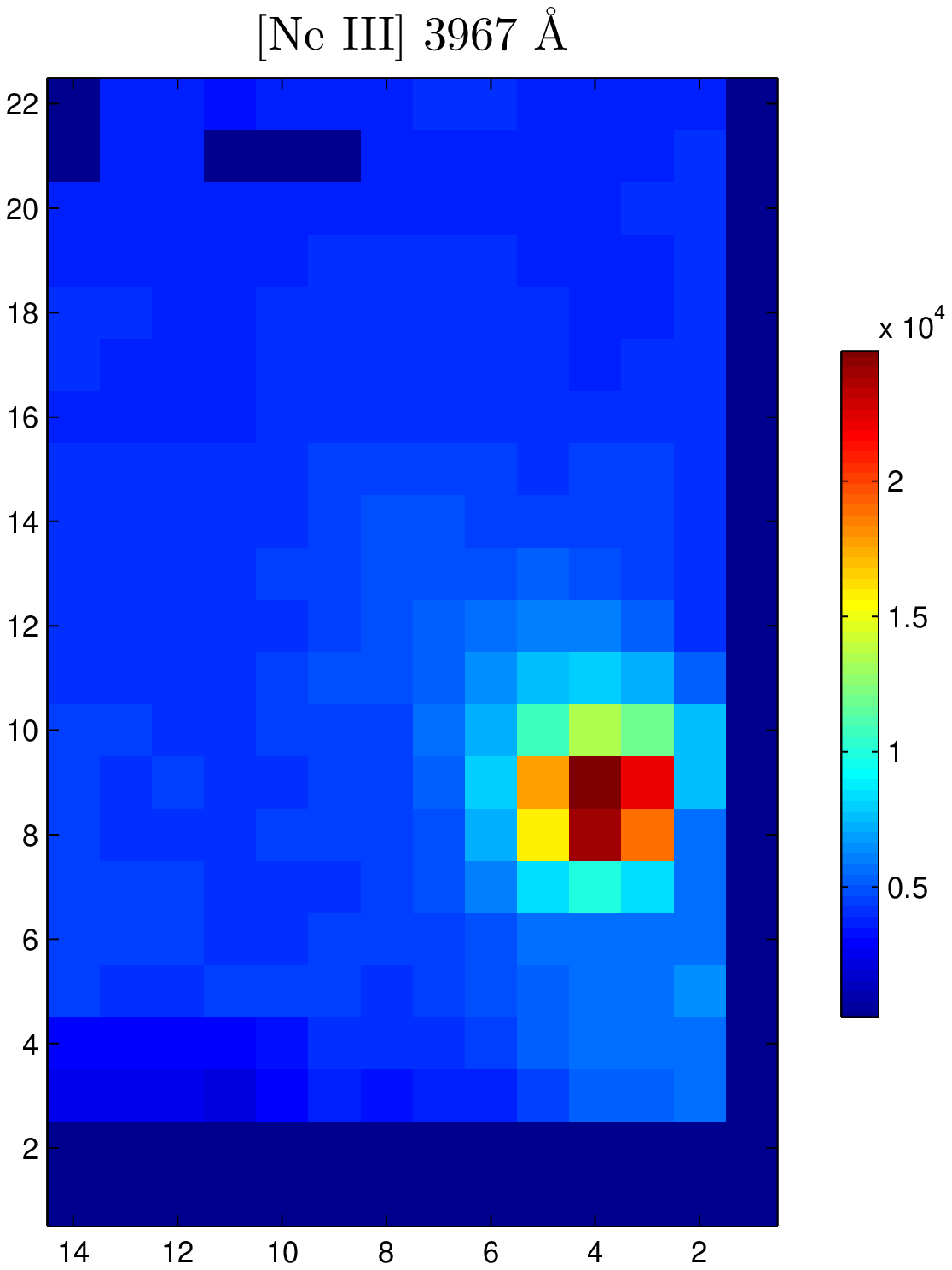, scale=0.42,
clip=, bbllx=170, bblly=175, bburx=500, bbury=675} \epsfig{file=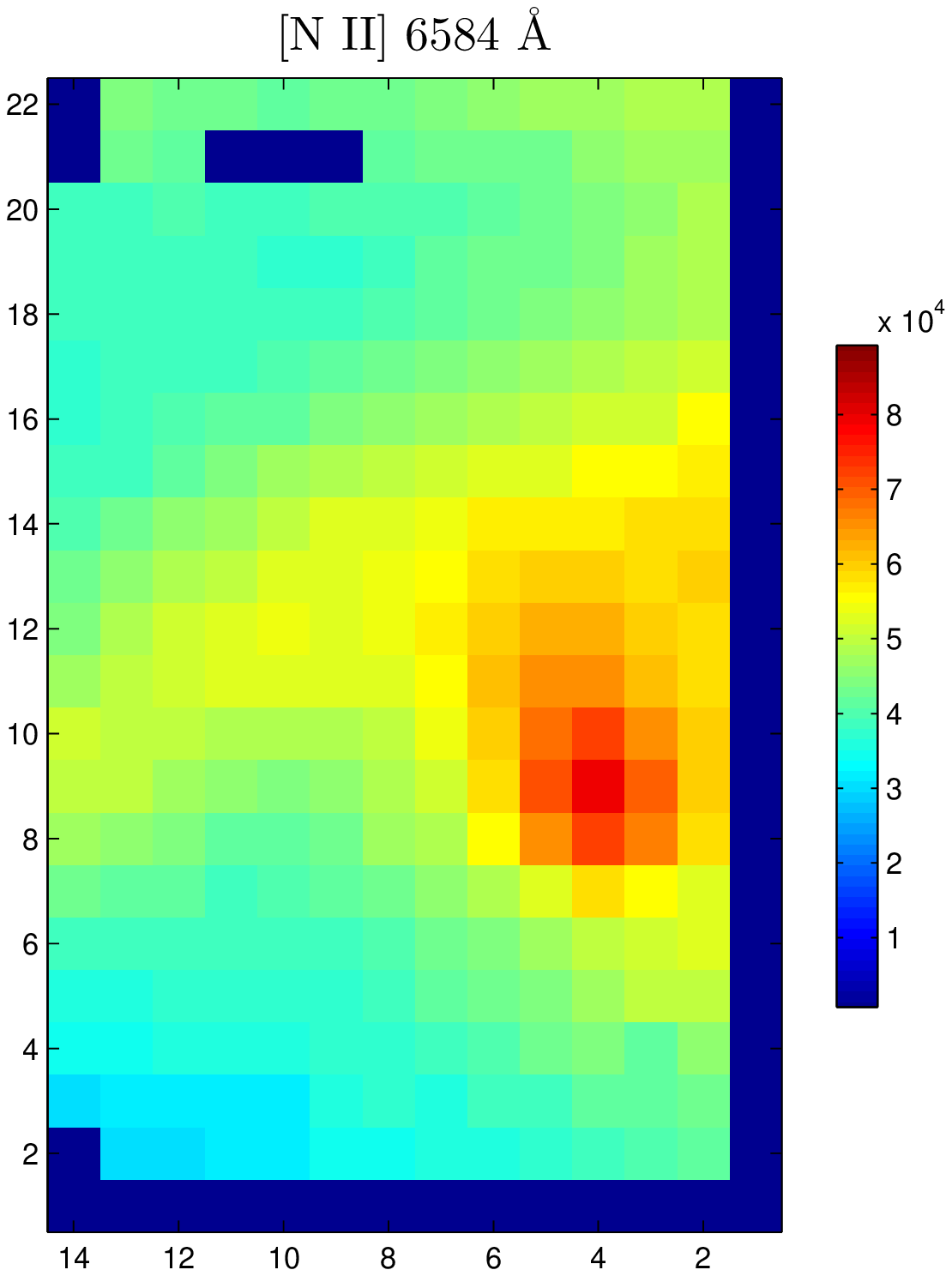,
scale=0.42, clip=, bbllx=170, bblly=175, bburx=500, bbury=675}
\epsfig{file=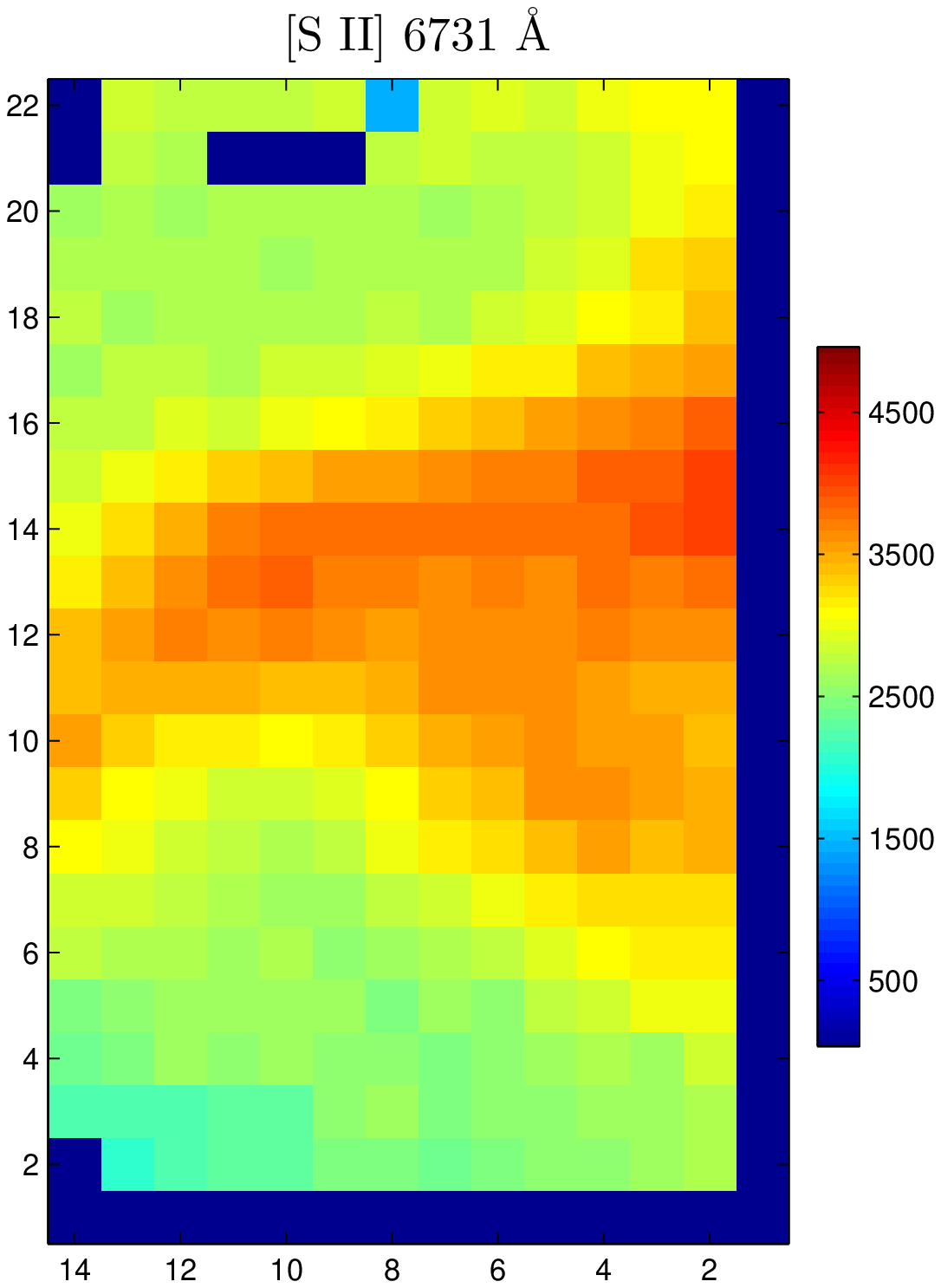, scale=0.42, clip=, bbllx=170, bblly=175, bburx=500,
bbury=675} \epsfig{file=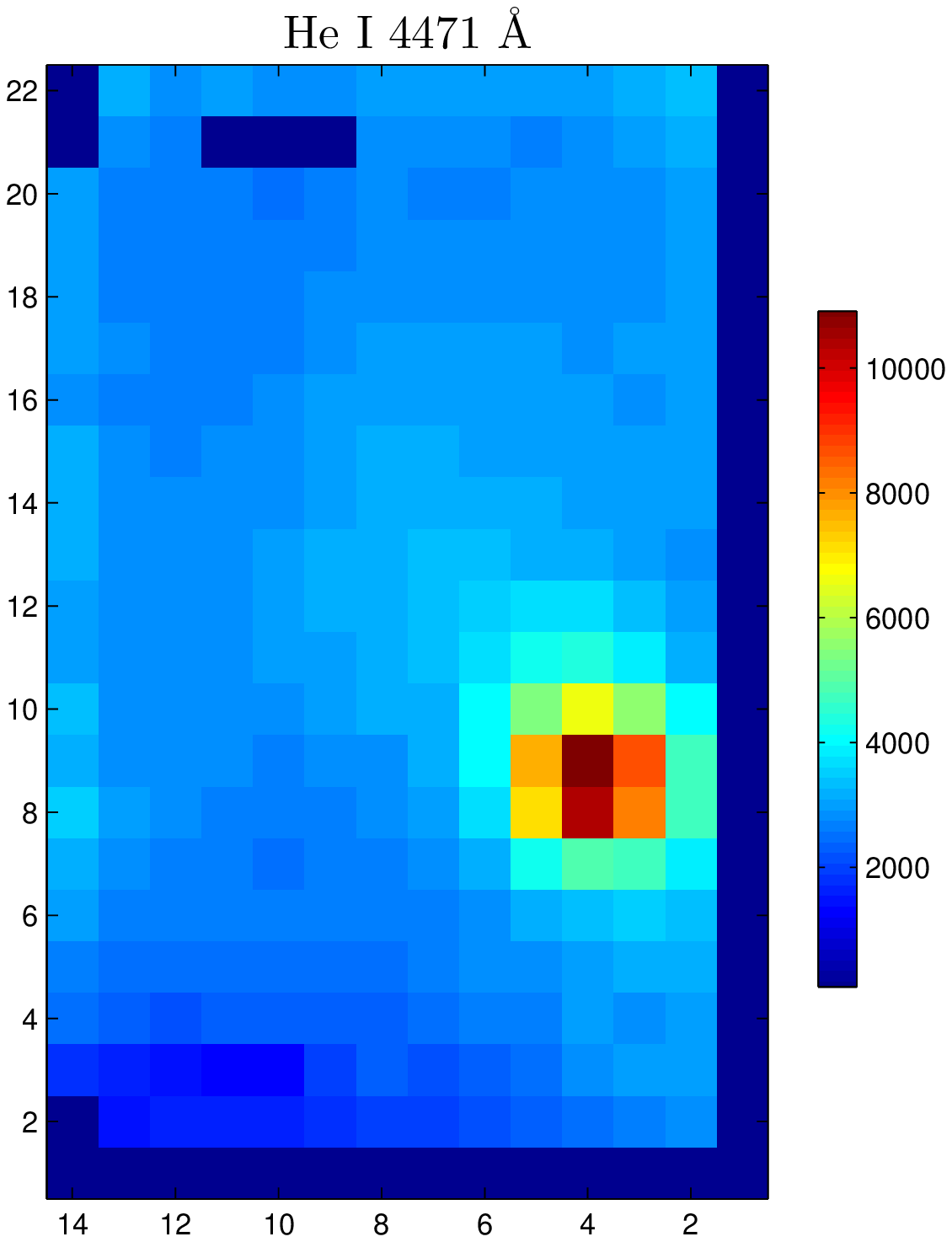, scale=0.42, clip=, bbllx=170, bblly=175,
bburx=500, bbury=675} \epsfig{file=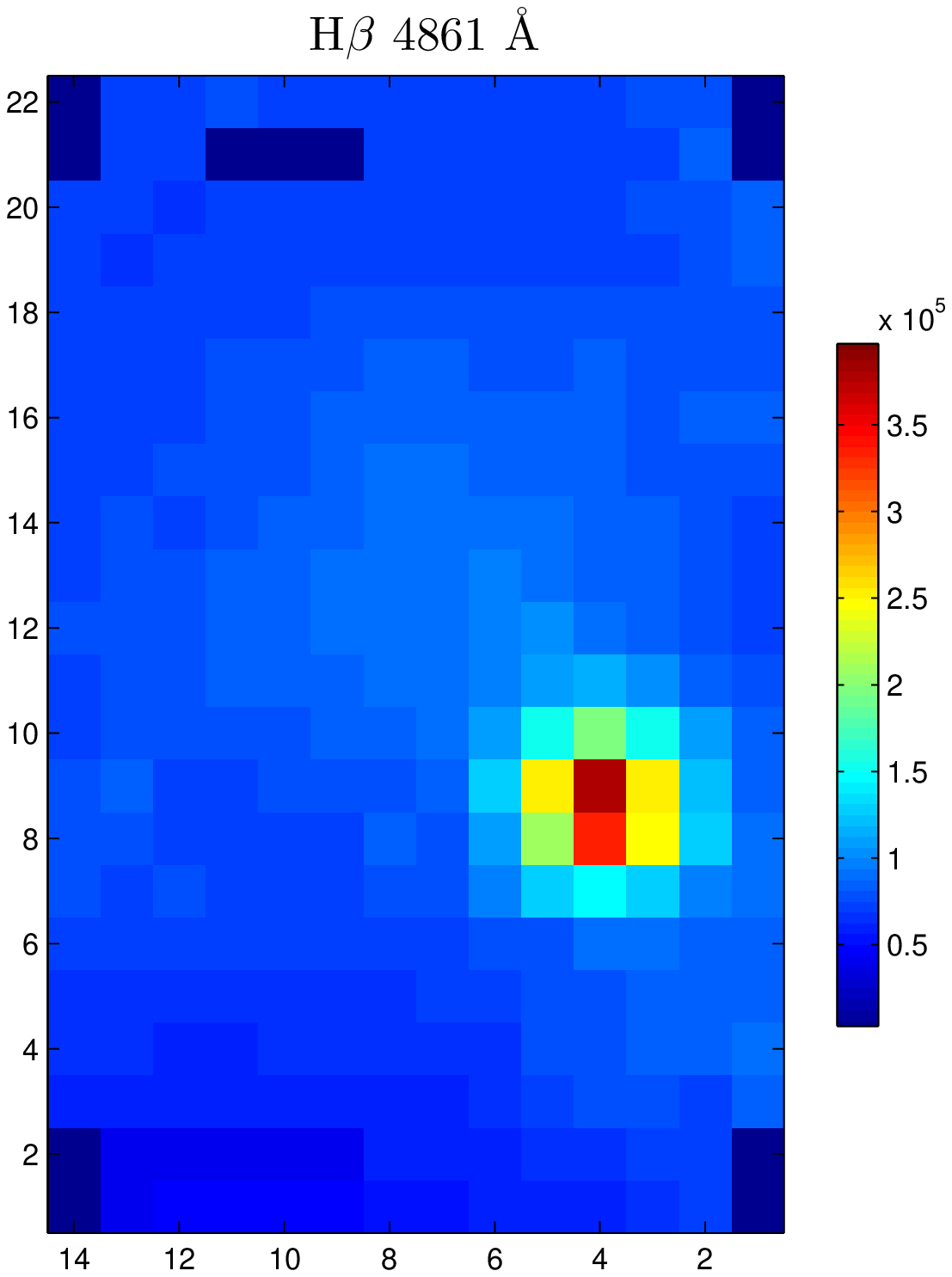, scale=0.42, clip=, bbllx=170,
bblly=175, bburx=500, bbury=675} \epsfig{file=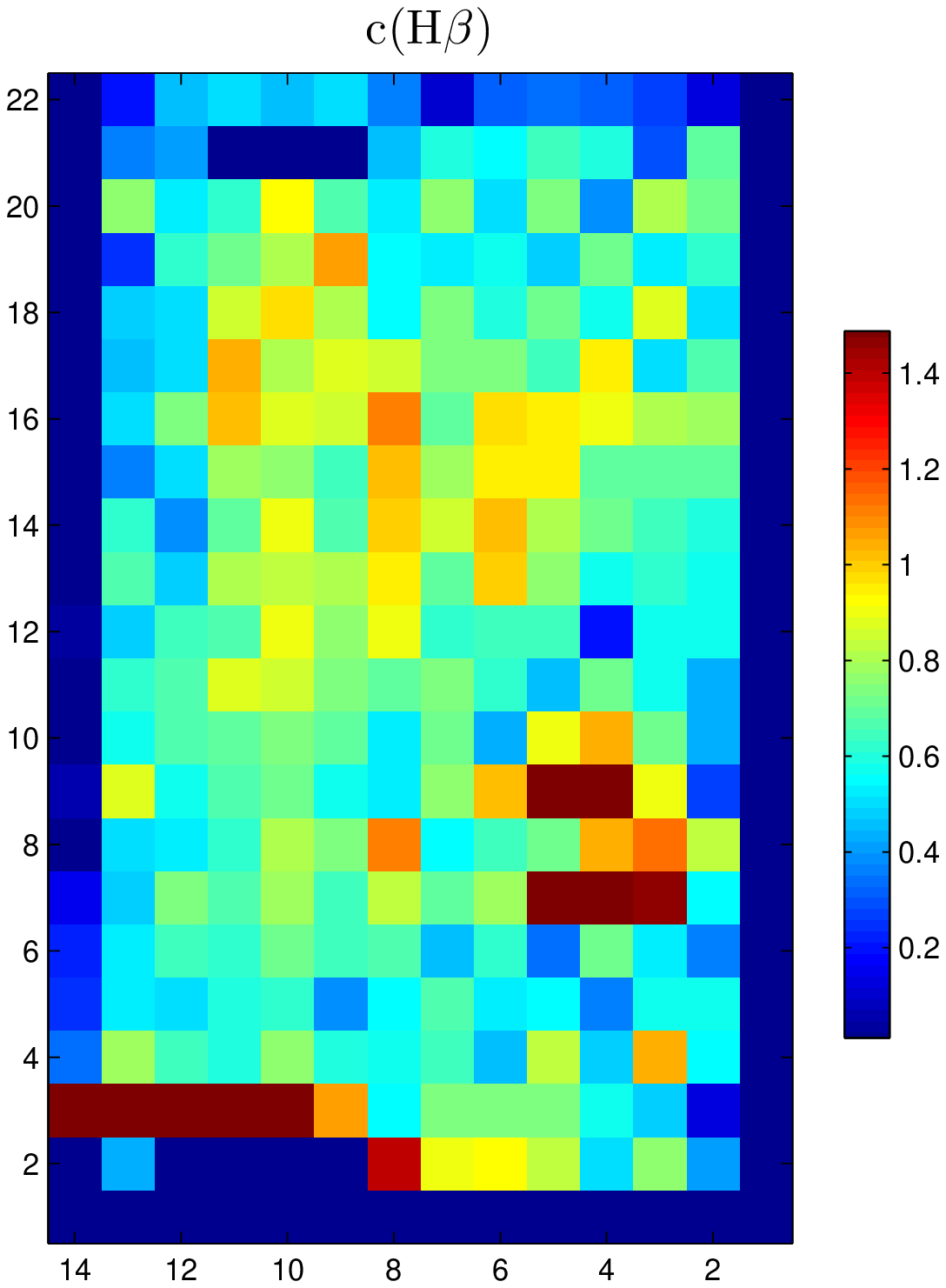, scale=0.42, clip=,
bbllx=170, bblly=175, bburx=500, bbury=675}

\caption{Monochromatic observed flux maps in emission lines arising from LV\,2
and vicinity (in units of 10$^{-18}$ erg s$^{-1}$ cm$^{-2}$ per spaxel).
Recorded with the 6.6$''$ $\times$ 4.2$''$ VLT Argus fibre array with
0.31$''$$\times$0.31$''$ spaxels (from top to bottom and left to right): \cii\
$\lambda$4267, \oii\ $\lambda$4649, \foiii\ $\lambda$5007, \fneiii\
$\lambda$3967, \fnii\ $\lambda$6584, \fsii\ $\lambda$6731, \hei\ $\lambda$4471,
\hb\ $\lambda$4861, and the reddening constant from the \hg/\hb\ ratio [the
spaxels which show up with large values on row 3 of the $c$(\hb) map have been
excluded from any analysis]. Spaxels (9--11, 21) correspond to broken fibres.
Dark blue masked out rows/columns are artifacts at the edges of the array
introduced during the correction for differential atmospheric refraction (2
spaxels per corner are blank by default as they correspond to sky fibres).}
\end{figure*}

\begin{figure}
\centering \epsfig{file=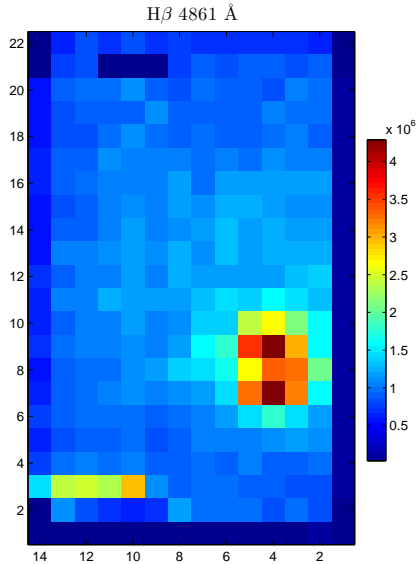, scale=0.5, clip=, bbllx=170, bblly=175,
bburx=500, bbury=675} \caption{A dereddened \hb\ image of LV\,2 computed using
the extinction map of Fig.\,~3 (in units of 10$^{-18}$ erg s$^{-1}$ cm$^{-2}$
per spaxel).}
\end{figure}

The FLAMES Argus data were used in two ways: to create spectral maps of LV\,2
and its vicinity, and to obtain summed spectra for various regions. The
($X$$=$14, $Y$$=$22, $\lambda$) spectral cube was co-added spatially over
spaxels (3--5, 7--10), (4--9, 11--15), and (4--11, 18--20) $+$ (7--12, 4--5)
$+$ (11--12, 6--17) to yield one-dimensional spectra for three regions
respectively encompassing the core, tail and local background of LV\,2 ($X$ and
$Y$ are respectively measured on the minor and major axis of the Argus array).
These regional spectra were scaled to the same number of spaxels and then the
background spectrum was subtracted from the other two so that pure proplyd
spectra were obtained. The subtraction resulted in spectra of LV\,2 that are
free from both nebular (M42) and telluric emission; these will be hereafter
called the Core and Tail spectra. To orientate the discussion, maps of LV\,2 in
several emission lines including \hb\ are shown in Fig\,~3. These show the
observed field without any background subtraction.

The proplyd is marginally resolved in the emission line maps, with a FWHM of
about 2.5 spaxels (0.78 arcsec). There is no conclusive evidence of this
varying between \hi\ and \hei, \oii, \cii, \foiii\ or \fneiii. However at the
longer wavelengths in the low ionization lines \fnii\ $\lambda$6584 and \fsii\
$\lambda$6731 the contrast of LV\,2 against the background is low and the FWHM
may be larger. In these lines there is a strong ionization front feature
extending across the field of view from NW to SE (seen also in Fig.\,~1); this
could arise in the nebula background but could also partially be intrinsic
emission from the tail itself. The form of the tail does not differ noticeably
between \hb, \hei, \foiii\ or \fneiii\ presenting a fan shape (as it does on
the \hst\ image of Fig.\,~1). The signal-to-noise in the \cii\ and \oii\ RLs
and \ffeiii\ is too low to say conclusively what the morphology of the tail is
in these species. The extinction map shown in Fig.\,~3 is not smooth on a
spaxel to spaxel basis and shows local maxima along the ridge of \fsii\
emission noted above and at the position of the proplyd core.

It is challenging to produce a complete VLT optical spectrum throughout the
range covered by the six Giraffe gratings taking into account that the LR1, 2,
3 set-ups were observed on a different night from the LR4, 5, 6 set-ups and
over slightly different atmospheric conditions. It was therefore deemed
appropriate to use the FOS `M42-background' observations as a benchmark to
which the Balmer decrement of the co-added LV\,2 background spectrum observed
by Argus (as defined above) was scaled. In this way the {\it mean} reddening
obtained for the LV\,2 background region is no longer a free parameter as it is
equalized to 0.7 (from Table 2, third column). To achieve this, scaling factors
of 1.070 and 0.833 were applied to the LR2 and LR6 spectra, respectively, while
the LR1 spectrum was scaled to the LR2 spectrum using the H~{\sc i}
$\lambda$3970 line common to both gratings. The LR3, 4, and 5 fluxes were left
unscaled. The same scaling factors were then applied to the co-added LV\,2 Core
and Tail spectra whose reddening constants are considered to be free
parameters.

\subsection{Nebular reddening}

Measured line fluxes and derived physical conditions are shown in Table 3 for
the ({\it background-subtracted}) Core and Tail, as well as the Background
spectra. The $c$(\hb) reddening coefficients were derived from a comparison of
the relative fluxes of the \ha, \hb, \hg, \hd, and H{$\epsilon$} lines with
their theoretical values from Storey \& Hummer (1995); \ha\ was saturated over
the Core and so the line was not considered for this position. The Blagrave et
al. (2007)--modified CCM law with $R_{\rm V}$ $=$ 5.5 was used as previously.
Resulting mean reddening constants are 1.20$\pm$0.08, 0.98$\pm$0.10 for the
LV\,2 Core and Tail regions (and 0.7 for the Background region). The $c$(\hb)
map computed from the \hg/\hb\ ratio if shown in Fig.\,~3. The reddening found
from the Argus data over LV\,2 Core is higher than the measurements by Blagrave
et al. (2007) who found $c$(\hb) $=$ 0.82$\pm$0.04 from \hst\ STIS data, but
for a position $\sim$30 arcsec to the south-west. At the same time, the
reddening derived by Blagrave et al. from FOS data is lower by 0.3 dex than
from STIS for almost identical positions. Both STIS and Argus sampled a larger
field of view than the FOS beam in each case and so the discrepancy may be due
to line of sight reddening variations over areas larger than the 0.$''$26 FOS
aperture. At a spatial resolution of 1.7 arcsec moderate values of $c$(\hb) $<$
1 were found in the M42 Trapezium region by O'Dell \& Yusef-Zadeh (2000). On
the other hand, using high resolution speckle interferometry Schertl et al.
(2003) found varying visual extinction of $A_V$ $\sim$ (1.9, 3.8), (0.7, 7.8,
$>$9.0), and 1.7 mag towards the resolved components of $\theta^1$ Ori A, B, C
respectively, or equivalently $c$ $\sim$ (0.9, 1.8), (0.3, 3.6, $>$4.2), and
0.8; they also note that stronger extinction is correlated with the presence of
circumstellar dust around the $\theta^1$ Ori A2 and B3 companions.

In conclusion, while systematic effects on the absolute value of $c$(\hb) are
difficult to determine, the variations shown on the extinction map of Fig.\,~3
and the peaking of c(\hb) over LV\,2, associated with the VLT spectra, are
suggestive of the presence of dust at that position. That the distribution of
dust particles in M42 is variable over small spatial scales was shown by the
mid-infrared mapping observations of Smith et al. (2005) at a
diffraction-limited resolution of 0.$''$35. Viewed in the context of their
figs\,~3 and 5, higher extinction values close to LV\,2 could be justified as
the proplyd and its immediate vicinity are a strong source of 11.7\,$\mu$m
continuum emission indicative of the presence of hot dust (100--150\,K).

\begin{figure}
\centering \epsfig{file=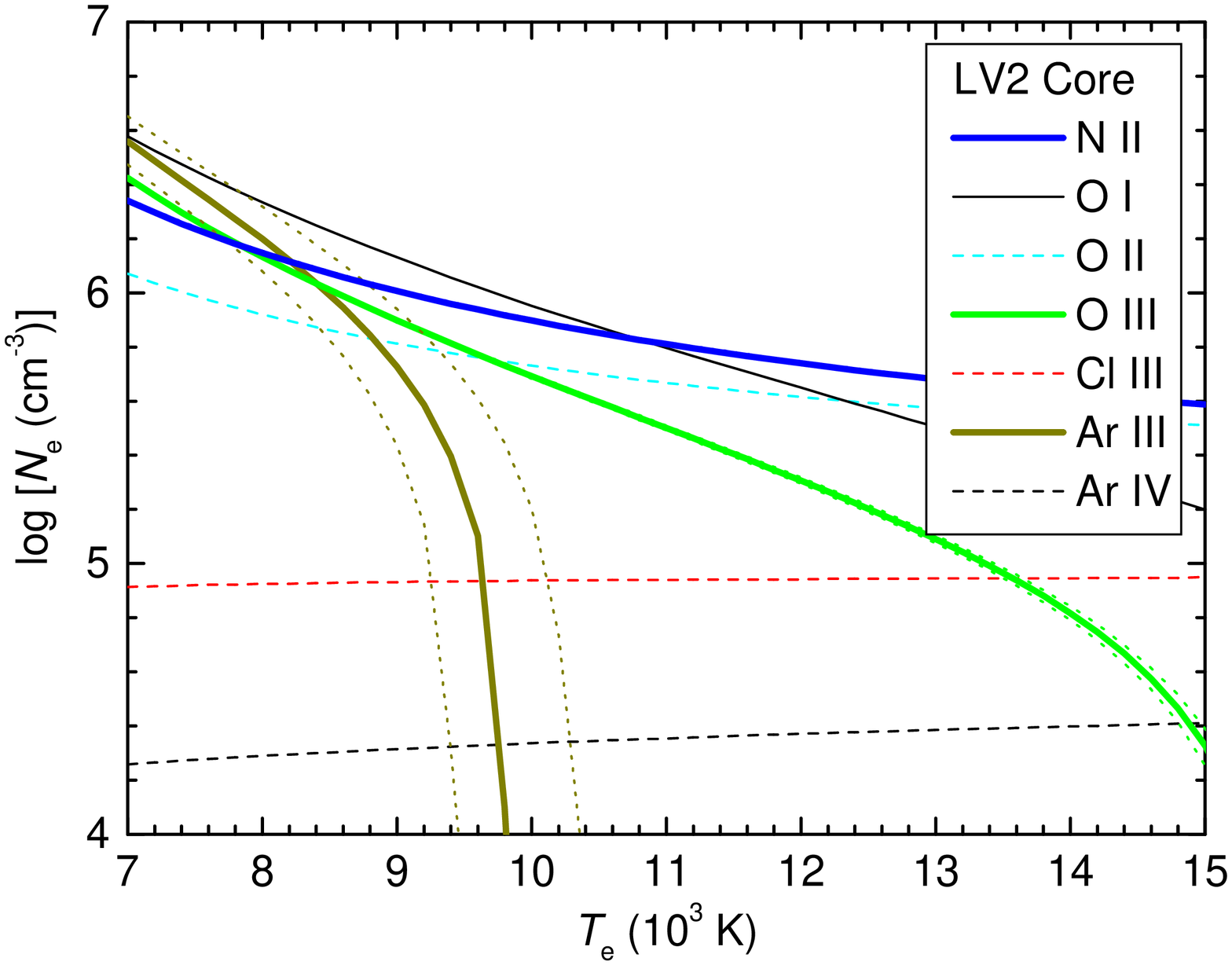, width=7 cm, scale=, clip=, angle=-90}
\caption{Electron temperature and density solutions for the
background-subtracted core of LV\,2 (as defined in the text). For \foiii\ and
\fariii\ dotted lines bracketing thick solid lines correspond to min/max values
of the diagnostic ratio. The \oii\ curve is from the \foii\
$\lambda$2470/($\lambda$3726+$\lambda$3729) \hst\ ratio.}
\end{figure}

\begin{figure*}
\centering \epsfig{file=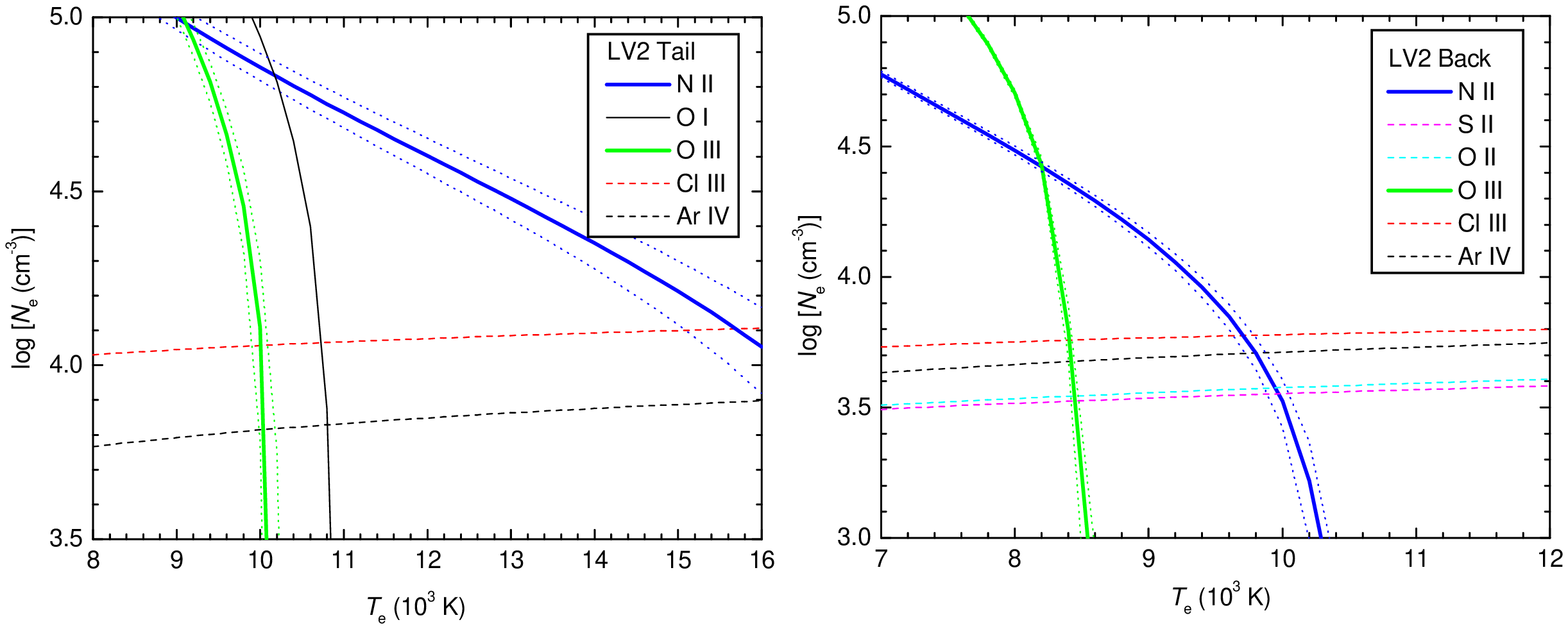, scale=0.6, clip=, bbllx=140,
bblly=80, bburx=470, bbury=790, angle=-90} \caption{Electron temperature and
density solutions for the (background-subtracted) tail and local background of
LV\,2. For \fnii\ and \foiii\ dotted lines bracketing thick solid lines
correspond to min/max values of the diagnostic ratio.}
\end{figure*}

\begin{figure*}
\centering \epsfig{file=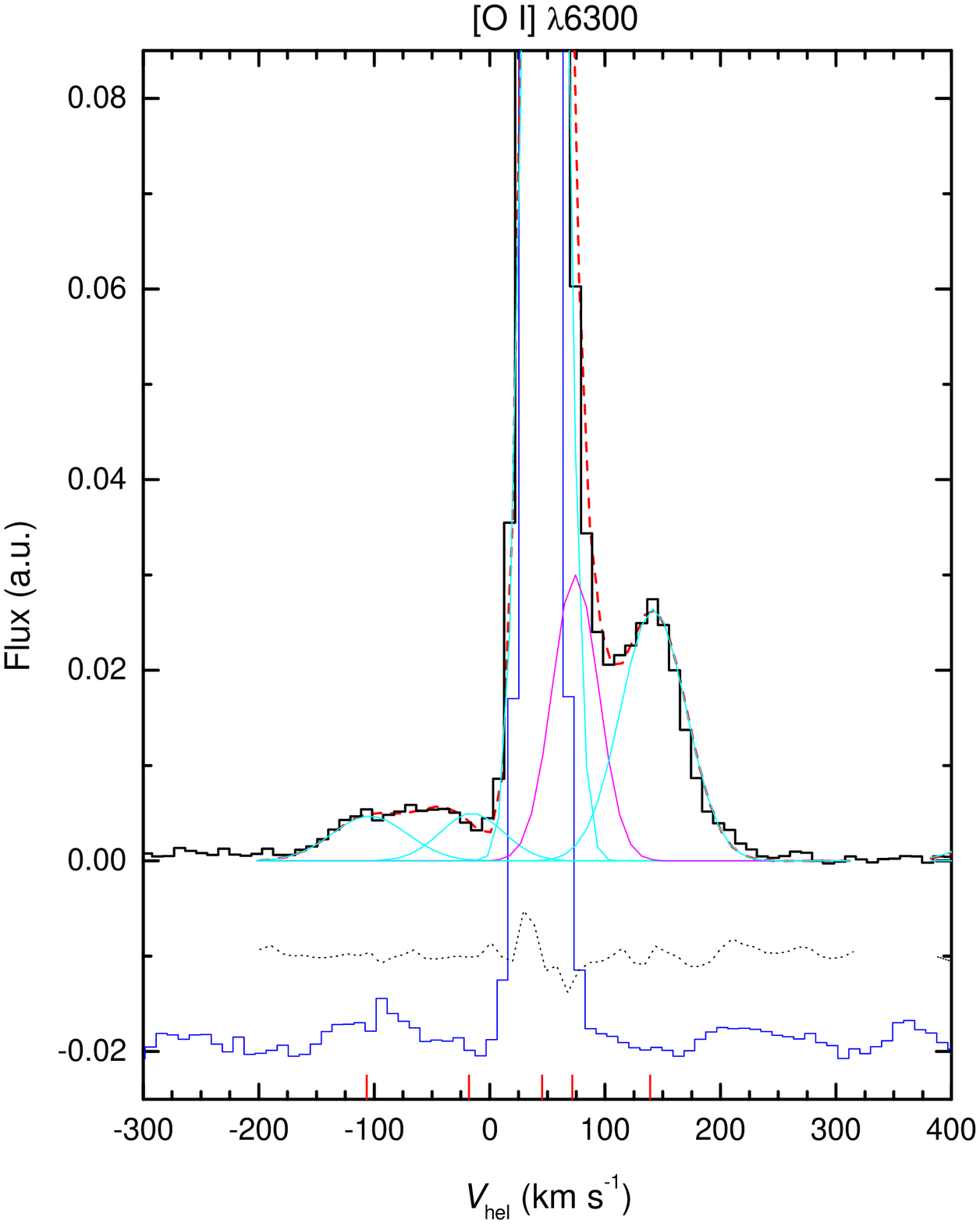, width=8 cm, scale=, clip=, angle=0}
\epsfig{file=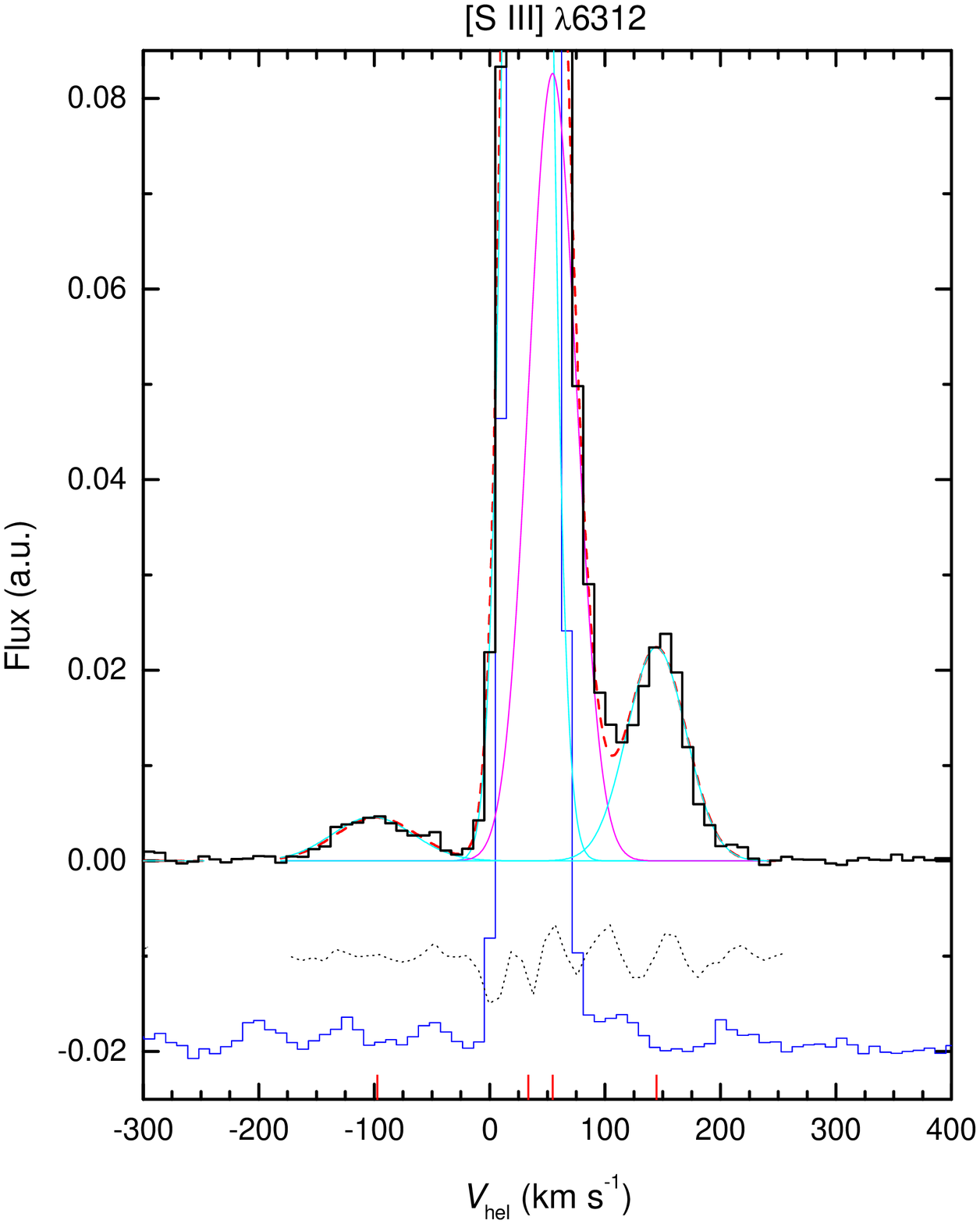, width=8 cm, scale=, clip=, angle=0} \caption{The
velocity profiles of \foi\ $\lambda$6300.34\,\AA\ and \fsiii\
$\lambda$6312.10\,\AA\ at a resolution of 9.5 km s$^{-1}$ per pixel from the
VLT Argus spectrum (LR5 grating). Black: the LV\,2 Core spectrum (observed
minus the Background). Dashed red: multiple Gaussian fit to the Core spectrum
-- the individual velocity components are shown in cyan/magenta, the red
vertical bars mark their centroids. The fit residuals are the dotted lines
(shifted vertically by $-$0.01). Blue: The LV\,2 Background spectrum as defined
in the text (multiplied by a factor of 3.5, so that the rest velocity peak
intensity of \foi\ matches the one in the Core spectrum, and shifted vertically
for clarity). }
\end{figure*}

\subsection{LV\,2 electron temperatures and densities}

\setcounter{table}{2}
\begin{table*}
\caption{Line fluxes and physical conditions of LV\,2 from VLT FLAMES
data.$^a$}
\begin{tabular}{lccc}
\noalign{\vskip3pt} \noalign{\hrule} \noalign{\vskip3pt}
                                              &Core       &Tail       &Background         \\

\noalign{\vskip3pt} \noalign{\hrule}\noalign{\vskip3pt} \noalign{\vskip3pt}

$c$(\hb)                                      &1.20$\pm$0.08     & 0.98$\pm$0.10         &0.70$\pm$0.03                               \\
$F$(\hb)                                    &(1.570$\pm$0.001)$\times$10$^{-13}$  &(1.820$\pm$0.001)$\times$10$^{-14}$     &(7.440$\pm$0.001)$\times$10$^{-14}$     \\
\noalign{\vskip2pt}
&\multicolumn{3}{c}{$I$($\lambda$)}\\

\foii\ $\lambda$3726                     & 9.46  :         & 53.0$\pm$2.0                &48.2$\pm$0.6                                 \\
\fneiii\ $\lambda$3868                   & 45.0$\pm$1.2    &30.7$\pm$1.6            &21.7$\pm$0.4                  \\
\fsii\ $\lambda$4069                     & 2.20$\pm$0.10     &2.90$\pm$0.08         &1.12$\pm$0.01         \\
\hei\ $\lambda$4471                      & 4.22$\pm$0.10     &4.14$\pm$0.10         &5.04$\pm$0.10          \\
\ffeiii\ $\lambda$4658                   & 0.049$\pm$0.008      &0.233$\pm$0.030    &0.395$\pm$0.001                 \\
\fariv\ $\lambda$4740                    & 0.262$\pm$0.006     &0.423$\pm$0.036     &0.273$\pm$0.004                                \\
\foiii\ $\lambda$4959                    & 119$\pm$0.4            & 111$\pm$0.6     &123$\pm$0.1                                  \\
\fariii\ $\lambda$5192                   & 0.151$\pm$0.004        & 0.089$\pm$0.015 & 0.059$\pm$0.024                     \\
\fcliii\ $\lambda$5539                   & 0.089$\pm$0.005        &0.399$\pm$0.025  &0.548$\pm$0.003                              \\
\hei\ $\lambda$5876                      & 11.8$\pm$0.3     &14.6$\pm$0.4           &17.6$\pm$0.1                \\
\foi\ $\lambda$6300                      & 1.54$\pm$0.05    &1.33$\pm$0.06          &0.751$\pm$0.010                                   \\
\fsiii\ $\lambda$6312                    & 2.18$\pm$0.07     &2.01$\pm$0.09         &2.02$\pm$0.02           \\
\fnii\ $\lambda$6584                     & 7.94$\pm$0.33     & 39.1$\pm$2.0         &32.0$\pm$0.5                                    \\
\hei\ $\lambda$6678                      & 1.94$\pm$0.09     &2.17$\pm$0.12         &3.53$\pm$0.06               \\
\fsii\ $\lambda$6731                     & 0.256$\pm$0.011   & 2.92$\pm$0.16        &2.89$\pm$0.04                                     \\
\noalign{\vskip2pt}
\cii\ $\lambda$4267                      & 0.223 $\pm$0.010   & 0.237$\pm$0.036      &0.218$\pm$0.005                     \\
\oii\ $\lambda$4638                      & 0.042 $\pm$0.007   &--                     &0.055$\pm$0.003         \\
\oii\ $\lambda$4641                      & 0.112 $\pm$0.008   &--                     &0.113$\pm$0.004                     \\
\oii\ $\lambda$4649                      & 0.154 $\pm$0.007   & 0.172$\pm$0.027      &0.163$\pm$0.004                       \\
\oii\ $\lambda$4650                      & 0.049 $\pm$0.007   &--                     &0.031$\pm$0.006                   \\
\oii\ $\lambda$4661                      & 0.043 $\pm$0.008   &--                     &0.052$\pm$0.003                   \\
\oii\ $\lambda$4676                      & 0.017 $\pm$0.006   &--                   &0.037$\pm$0.004   \\
&\multicolumn{3}{c}{$I$($\lambda$) ratios}\\
\foiii\ $\lambda$4959/$\lambda$4363           & 16.4$\pm$0.2     &50.7$\pm$1.6          &98.4$\pm$0.7                        \\
\foi\ $\lambda$6300/$\lambda$5577             & 23.3$\pm$1.4     & 63.1:                 & --                               \\
\fnii\ $\lambda$6584/$\lambda$5755            & 2.78$\pm$0.02    & 21.1$\pm$1.3          &61.2 $\pm$1.2                                  \\
\fsii\ $\lambda$6731/$\lambda$6716            & 2.64$\pm$0.01    & 2.35$\pm$0.04         &1.66$\pm$0.004                       \\
\foii\ $\lambda$3726/$\lambda$3729            & 2.50:            & 2.88$\pm$0.08         &1.91$\pm$0.003                       \\
\fcliii\ $\lambda$5539/$\lambda$5518          & 3.87$\pm$0.02    & 2.02$\pm$0.27         &1.52$\pm$0.02              \\
\fariv\ $\lambda$4740/$\lambda$4711           & 2.30$\pm$0.02    & 1.26$\pm$0.18         &1.15$\pm$0.03                       \\

\noalign{\vskip2pt}
&\multicolumn{3}{c}{\eld\ (\cmt)}\\
 \foi\                                 & 1.4$^{+0.1}_{-0.2}$$\times$10$^{6}$      &--                                       &-- \\
 \fsii\                                & $\gtrsim$10$^4$                          &$\gtrsim$10$^4$              &3400$\pm$200       \\
 \foii\                                & 13300:                                   &$\gtrsim$10$^4$                        &3500$\pm$200       \\
 \fcliii\                              & 8.3$^{+0.3}_{-0.3}$$\times$10$^{4}$      &1.1$^{+0.4}_{-0.3}$$\times$10$^{4}$    &5800$\pm$900       \\
 \fariv\                               & 2.1$^{+0.1}_{-0.1}$$\times$10$^{4}$      &6500$\pm$2000                          &4800$\pm$1000       \\
\fnii\                                 & 1.0$^{+0.2}_{-0.1}$$\times$10$^{6}$      &8.0$^{+0.1}_{-0.1}$$\times$10$^{4}$                      & --        \\
&\multicolumn{3}{c}{\elt\ (K)}\\
 \foiii\                               & 9000$\pm$600                               &9700$\pm$300   &8400$\pm$200     \\
 \fnii\                                & --                                         & --            &9700$\pm$500       \\

\noalign{\vskip3pt} \noalign{\hrule}\noalign{\vskip3pt}
\end{tabular}
\begin{description}
\item[$^a$] $F$(\hb) denotes observed fluxes in erg\,cm$^{-2}$\,s$^{-1}$ per
0.31$''$$\times$0.31$''$ spaxel; $I$($\lambda$) denotes dereddened relative
intensities in units of \hb\ $=$ 100. The Core and Tail values refer to
background-subtracted spectra. All entries are for the rest velocity components
of the lines. Entries referring to \foii\ $\lambda\lambda$3726, 3729 in the
Core spectrum are uncertain as the peak of LV\,2 was very close to the edge of
the field of view at this wavelength. `:' denotes very uncertain value.

\end{description}
\end{table*}

Electron densities and temperatures from the Argus data were derived from
various diagnostics. In Figs\,~5 and 6 a group of curves is shown on the (\elt,
\eld) diagnostic plane representing solutions corresponding to the observed
line ratios from Table~3. It should be noted that these ratios are for the rest
velocity components of the lines in each case. For the Background region \elt\
$=$ 9700$\pm$500\,K is obtained for the singly ionized species and
8400$\pm$200\,K for all other species with \eld\ $=$ 4700$\pm$1400\,\cmt. For
the Tail region \elt\ $=$ 9700$\pm$300\,K and log[\eld\ (\cmt)] $=$
4.58$^{+0.24}_{-0.58}$ is adopted for all species. For the Core region, which
is expected to be very dense, both the \foiii\ and especially the \fnii\
nebular to auroral ratios become more sensitive probes of \eld\ than they are
of \elt; hence, the \fnii\ curve indicates \eld $>$ 6.7$\times$10$^5$ \cmt\ for
\elt\ $<$ 10\,000\,K and the intersection with the \foiii\ and \fariii\
$\lambda$5192/$\lambda$7135 curves yields (\elt, \eld) $\approx$ (8400\,K,
1.1$\times$10$^6$ \cmt). At such high densities the usual optical diagnostic
ratios -- \foii\ $\lambda$3726/$\lambda$3729, \fsii\
$\lambda$6716/$\lambda$6731, \fcliii\ $\lambda$5518/$\lambda$5538, \fariv\
$\lambda$4711/$\lambda$4740 -- do not return reliable measurements as both
their constituent lines are equally affected by collisional de-excitation. Here
for the Core we adopt representative values of \elt\ $=$ 9000 $\pm$ 600\,K and
log[\eld\ (\cmt)] $=$ 5.90$^{+0.10}_{-0.14}$ for all ions for which abundances
are derived in Section~5. The upper density limit is consistent with the
\fciii\ $\lambda$1907/$\lambda$1909 ratio observations by Henney et al. (2002).

Additional evidence for the existence of high density ionized gas associated
with LV\,2 is given by the analysis of several \ffeiii\ 3d$^6$ lines detected
in the LR3 set-up. The rest velocity components of these lines are weak in the
Core spectrum of LV\,2 (only the strongest line in the multiplet
$\lambda$4658.10 is detected along with $\lambda$4881), but their redshifted
counterparts arising in gas outflowing from the Core are well-detected. This
indicates that the abundance of iron must be enhanced in the redshifted outflow
with respect to the Core and to the mean M42 value. An intensity-weighted mean
of nine lines yields a heliocentric velocity of $+$155 $\pm$ 3 km s$^{-1}$ for
their emitting region. The $\lambda\lambda$4607.0, 4881.1, 4931.0, and 5011.3
redshifted components show intensity ratios, relative to \ffeiii\
$\lambda$4658.10, of 0.103, 0.075, 0.153 and 0.382 respectively; a comparison
with the theoretical intensities of Keenan et al. (2001) yields densities of
(1.0 to 3.8)$\times$10$^6$ \cmt. The above lines return unique solutions to the
density. Several more lines are detected ($\lambda\lambda$4701.62, 4733.90,
4769.40, 4777.70\,\AA) which provide both low and high solutions to the
density. In all these cases, however, the high density solution is (2 to
4)$\times$10$^6$ \cmt\ in agreement with the unique values found above.
Arguably, if the outflow is this dense then the zero velocity ionized component
at LV\,2 could be denser.

Thus both the diagnostic ratio diagrams and the \ffeiii\ lines indicate that
the ionized gas in the Core of LV\,2 and in the proplyd redshifted outflow is
of high density. An independent confirmation of our analysis is provided by
Henney et al. (2002) who derived \eld\ $=$ 10$^6$ \cmt\ for the bright cusp of
the proplyd from the \fciii\ $\lambda$1907/$\lambda$1909 \hst\ STIS ratio. (The
individual \fciii\ doublet components are not resolved in the FOS spectrum of
Fig.\,~2).

A measurement of the neutral gas density in LV\,2 is possible using the \foi\
$\lambda$6300/$\lambda$5577 ratio which is sensitive to densities up to
$\sim$10$^8$ \cmt. The $\lambda$6300.34 line shows several velocity components
in the Argus LR5 grating spectrum of the Core. A multiple Gaussian fit to the
line profile is shown in Fig\,~7. Up to five components can be fitted. Two
redshifted components are seen, each at a flux level of about 20 per cent with
respect to the rest velocity component (Table~4). The blueshifted components
have a total flux of 9 per cent with respect to the rest velocity. In the
Background spectrum only the 0 km s$^{-1}$ component is seen, and hence we
conclude that the negative and positive velocity counterparts to the line are
associated with LV\,2 and arise from neutral gas present in the bipolar jet of
the proplyd.

The \foi\ $\lambda$5577 line in the LR4 grating shows only the two redshifted
components at $V_{\bigodot}$ $=$ $+$62.7 $\pm$ 1.6 and $+$101 $\pm$ 1 km
s$^{-1}$ with a total flux of 20 percent with respect to the rest component;
these are again absent from the Background. We thus took two \foi\
$\lambda$6300/$\lambda$5577 ratios using first just the summed flux of the
redshifted components and then the rest velocity components: these are 37.2 and
23.3 respectively (dereddened). At 8400\,K they return \eld's of
9.5$\times$10$^5$ and 1.8 $\times$10$^6$ \cmt\ for the redshifted jet and the
Core of LV\,2 respectively. These results show that there is a reservoir of
dense neutral gas in the proplyd which is being photoevaporated from it
becoming entrained in a bipolar outflow. \footnote{A measurement of the density
using \foi\ in the Background spectrum is not possible as the $\lambda$5577
line is abnormally strong compared to published values for M42 from Esteban et
al. (2004) -- this indicates that the line suffers from telluric contamination:
the Core and Tail spectra are immune to this as they are
background-subtracted.}  This direct measurement of the density in the partly
neutral outflow of LV\,2 has repercussions for the mass-loss rate from the
proplyd.

The fact that very high \emph{electron} densities are probed via a neutral gas
proxy (the \foi\ ratio) leads to the conclusion that the emitting O$^0$ volume
itself is substantially ionized (while at the same time the density of the
inner neutral core of LV\,2 must be even larger depending on its temperature).
This is supported by the fact that emission from the highly ionized species
\fsiii\ is seen at similar velocities.  An examination of the \fsiii\
$\lambda$6312.10 line shows that it has one blueshifted and two redshifted
velocity components present only in the LV\,2 Core spectrum (Fig.\,~7 and
Table~4). Depending on the exact geometry of the outflow this picture would
suggest that the gas escaping from LV\,2 becomes almost instantly ionized.

\setcounter{table}{3}
\begin{table}
\caption{The \foi\ 6300.34, \fsiii\ 6312.10, and \ffeiii\ 4658.10\,\AA\ line
structure in the intrinsic LV\,2 spectrum.$^a$ }
\begin{tabular}{rcc}
\noalign{\vskip3pt} \noalign{\hrule} \noalign{\vskip3pt}
$V_{\bigodot}$ (km\,s$^{-1}$)       &FWHM (km\,s$^{-1}$)   &Relative flux       \\
&\multicolumn{2}{c}{\foi}              \\
$-$107$\pm$12    &    71.6$\pm$18.9      & 0.05$\pm$0.01                 \\
$-$39.3$\pm$8.9  &   57.7$\pm$13.9       & 0.04$\pm$0.01                  \\
44.9$\pm$0.3     &  18.5$\pm$1.1         & 1.00$\pm$0.06                  \\
71.1$\pm$7.7     &  38.0$\pm$10.4        & 0.18$\pm$0.06                   \\
139$\pm$1        &   60.6$\pm$2.7        & 0.22$\pm$0.02                    \\
&\multicolumn{2}{c}{\fsiii}                \\
$-$97.7$\pm$6.2       & 77.8$\pm$15.0       & 0.04$\pm$0.01                 \\
33.0$\pm$0.4        & 21.1$\pm$1.6        & 1.00$\pm$0.15                 \\
54.4$\pm$7.5         & 40.0$\pm$5.7        & 0.43$\pm$0.16                  \\
144$\pm$1           & 53.5$\pm$3.2        & 0.14$\pm$0.02                   \\
&\multicolumn{2}{c}{\ffeiii}                \\
$-$97.1$\pm$1.9       & 54.0$\pm$4.9      & 3.30$\pm$0.76                 \\
31.7$\pm$1.9          & $<$16.7       & 1.00$\pm$0.20                 \\
155$\pm$1             & 35.9$\pm$1.3       & 11.1$\pm$2.3                   \\
\noalign{\vskip3pt} \noalign{\hrule}\noalign{\vskip3pt}
\end{tabular}
\begin{description}
\item[$^a$] Velocities are heliocentric. The FWHMs have been corrected for
instrumental broadening which for the LR5 grating is 28$\pm$1 km\,s$^{-1}$ from
Th-Ar arc line measurements. Fluxes are relative to the rest velocity
component.
\end{description}
\end{table}

\setcounter{table}{4}
\begin{table*}
\caption{Ionic abundances in LV\,2 and its M42 vicinity (in a scale where
log\,H $=$ 12). The employed metallic lines were CELs but RL abundances are
also listed for \cpp\ and \opp.}
\begin{tabular}{lcccccc}
\noalign{\vskip3pt} \noalign{\hrule} \noalign{\vskip3pt}
                        & Core$^a$   & Tail$^b$   & Background$^c$   &Core$^d$          &Tail$^e$       &M42$^f$ \\
                        &   VLT      &    VLT     &       VLT        &  FOS         &  FOS              &FOS        \\
\noalign{\vskip3pt} \noalign{\hrule}\noalign{\vskip3pt}
\noalign{\vskip3pt}
\cp\                         &   --         & --            & --             & 8.40$\pm$0.17   &6.35$\pm$0.11:   &  8.17$\pm$0.17        \\
\cpp\                        &   --         & --            & --             & 8.84$\pm$0.19   &--               &  8.02$\pm$0.19     \\
\cpp\ (RL)                   &8.32$\pm$0.05 &8.35$\pm$0.05  &8.32$\pm$0.05   & --              &--               &-- \\
\np\                         &7.44$\pm$0.14 &7.17$\pm$0.10  &6.91$\pm$0.10   & 7.68$\pm$0.14   &7.29$\pm$0.10    &  7.66$\pm$0.05     \\
\op\                         &  --          &8.30$\pm$0.23  &7.90$\pm$0.17   & 8.75$\pm$0.18   &7.90$\pm$0.09    &  8.53$\pm$0.07     \\
\opp\                        &8.63$\pm$0.14 &8.13$\pm$0.06  &8.38$\pm$0.04   & 8.71$\pm$0.15   &8.32$\pm$0.10    &  8.08$\pm$0.04     \\
\opp\ (RL)                   &8.54$\pm$0.10 &8.54$\pm$0.10  &8.57$\pm$0.10   &--               &--               &-- \\
\nepp\                       &7.94$\pm$0.12 &7.58$\pm$0.05  &7.72$\pm$0.05   & 7.96$\pm$0.12   &7.74$\pm$0.12    &  7.38$\pm$0.06     \\
\sulp\                       &5.73$\pm$0.12 &5.75$\pm$0.13  &5.37$\pm$0.12   & --              &5.51$\pm$0.08    &  6.15$\pm$0.12     \\
\sulpp\                      &6.77$\pm$0.10 &6.67$\pm$0.03  &7.03$\pm$0.05   & 6.63$\pm$0.11   &7.07$\pm$0.11    &  6.60$\pm$0.05     \\
\clpp\                       &5.18$\pm$0.16 &4.83$\pm$0.11  &5.07$\pm$0.05   & --              &  --             &  --              \\
\arpp\                       &6.44$\pm$0.12 &6.04$\pm$0.05  &6.27$\pm$0.05   & 6.53$\pm$0.08   &6.27$\pm$0.09    &  6.03$\pm$0.11     \\
\arppp\                      &5.30$\pm$0.14 &5.05$\pm$0.05  &5.17$\pm$0.05   & --              &--               &  --                     \\
\fepp\                       &4.67$\pm$0.10 &5.16$\pm$0.07  &5.56$\pm$0.05   & --              &--               &  --                     \\
\noalign{\vskip3pt} \noalign{\hrule}\noalign{\vskip3pt}
\end{tabular}
\begin{description}

\item[$^a$] The core spectrum is background subtracted. \elt\ $=$ 9000 $\pm$
600\,K and log[\eld\ (\cmt)] $=$ 5.90$^{+0.10}_{-0.14}$ was adopted for all
ions.

\item[$^b$] The tail spectrum is background subtracted. Adopted \elt\ $=$ 9700
$\pm$ 300\,K and log[\eld\ (\cmt)] $=$ 4.58$^{+0.24}_{-0.58}$.

\item[$^c$] Adopted \elt\ $=$ 8400 $\pm$ 200\,K for all non-singly ionized
species and 9700 $\pm$ 500\,K for singly ionized species with \eld $=$ 4700
$\pm$ 1400 \cmt.

\item[$^d$] The core spectrum is background subtracted. Adopted \elt\ $=$ 9050
$\pm$ 650\,K with log[\eld\ (\cmt)] $=$ 5.92$^{+0.08}_{-0.10}$. The entries for
the \cpp\ and \opp\ RL abundances correspond to those of Core (VLT) as the FOS
observation is spatially contained within the Core (VLT) observation.

\item[$^e$] Adopted \elt\ $=$ 8950$^{+1550}_{-1250}$\,K with log[\eld\ (\cmt)]
$=$ 4.00$^{+0.90}_{-0.32}$. The \cp\ entry is very uncertain as it was obtained
from a tentative \cii] $\lambda$2326 detection.

\item[$^f$] Adopted \elt\ $=$ 8850$\pm$250\,K with log[\eld\ (\cmt)] $=$
4.61$^{+0.20}_{-0.38}$.
\end{description}
\end{table*}

\section{Chemical abundances}

\setcounter{table}{5}
\begin{table*}
\caption{The helium abundance in LV\,2 and M42 vicinity.$^a$}
\begin{tabular}{lcccc}
\noalign{\vskip3pt} \noalign{\hrule} \noalign{\vskip3pt}
                          &Core                          &Tail               &Background              &M42 filament       \\            
                          & FOS                          &VLT               & VLT                     &FOS                      \\     
\noalign{\vskip3pt} \noalign{\hrule}\noalign{\vskip3pt}

\noalign{\vskip3pt}
\hep\ ($\lambda$4471)       &0.086 $\pm$0.018           &0.078 $\pm$0.015   &0.097 $\pm$0.015         &0.071$\pm$0.016                \\           
\hep\ ($\lambda$5876)       &0.094 $\pm$0.009           &0.095 $\pm$0.010   &0.118 $\pm$0.011         &0.071$\pm$0.011                \\           
\hep\ ($\lambda$6678)       &0.102 $\pm$0.038           &0.053 $\pm$0.035   &0.086 $\pm$0.029         &0.078$\pm$0.042                \\           
\noalign{\vskip2pt}
\hep/\hp\ avg.              &0.094 $\pm$0.014           &0.083 $\pm$0.010   &0.108 $\pm$0.010         &0.073$\pm$0.015                         \\
He/H                        &0.104$\pm$0.018            &0.088$\pm$0.012    &0.151 $\pm$0.039        &0.094$\pm$0.026                        \\

\noalign{\vskip3pt} \noalign{\hrule}\noalign{\vskip3pt}
\end{tabular}
\begin{description}
\item[$^a$] The core and tail values are from background-subtracted spectra.
Individual values were given weights of 1:3:1 in the computation of the mean
according to the relative line intensities. The total He/H ratios incorporate
corrections for the presence of neutral helium (Eq.\,~2).

\end{description}
\end{table*}

\setcounter{table}{6}
\begin{table*}
\caption{Total gas-phase abundances for the background-subtracted core and tail
of LV\,2, the local background, and a M42 filament (adopting $t^{2}$ $=$ 0).
Independent values for M42 (gas-phase) and the solar photosphere are listed (in
units where log\,H $=$ 12).}
\begin{tabular}{lccccccc}
\noalign{\vskip3pt} \noalign{\hrule} \noalign{\vskip3pt}
                        & LV\,2 Core       & LV\,2 Tail       & LV\,2 Background               & Filament            &M42 ($t^2$ $=$ 0)    &M42 ($t^2$ $=$ 0.022)  &Sun         \\
Element                 & [1]              & [1]              & [1]                            & [1]                 &[2]                   & [2]                   & [3]   \\
\noalign{\vskip3pt} \noalign{\hrule}\noalign{\vskip3pt}
He                      & 10.973          & 10.919           &  11.033                            & 10.863           &--                 & 10.942                  & 10.93$\pm$0.01                                             \\
He                      & 11.017          & 10.944           &  11.179                            & 10.973           &10.991             & 10.988                  &  --                                       \\
C                       & 8.98$\pm$0.15   &   --             &  (8.40$\pm$0.13)                   & 8.40$\pm$0.13       &--                 & --                   & 8.43$\pm$0.05                                                       \\
C (RL)                  & 8.66$\pm$0.10   &  8.35$\pm$0.11:  &  8.55$\pm$0.08                     & --               &8.42$\pm$0.02      & 8.42$\pm$0.02           &  --                            \\
N                       & 7.86$\pm$0.17   &  7.43$\pm$0.12   &  7.46$\pm$0.15                     & 7.79$\pm$0.13       &7.65$\pm$0.09      & 7.73$\pm$0.09        & 7.83$\pm$0.05                                         \\
O                       & 9.03$\pm$0.13   &  8.53$\pm$0.15   &  8.51$\pm$0.05                     & 8.66$\pm$0.08       &8.51$\pm$0.03      & 8.67$\pm$0.04        & 8.69$\pm$0.05                                              \\
O (RL)                  & 8.96$\pm$0.10   &  8.74$\pm$0.10   &  8.65$\pm$0.05                     & --               &8.63$\pm$0.03      & 8.65$\pm$0.03           & --                                               \\
Ne                      & 8.28$\pm$0.06   &  7.93$\pm$0.15   &  7.86$\pm$0.07                     & 7.97$\pm$0.10       &7.78$\pm$0.07      & 8.05$\pm$0.07        & 7.93$\pm$0.10                                               \\
S                       & 6.83$\pm$0.25   &  6.74$\pm$0.25   &  7.04$\pm$0.13                     & 6.80$\pm$0.15       &7.06$\pm$0.04      & 7.22$\pm$0.04        & 7.12$\pm$0.03                                           \\
Cl                      & 5.36$\pm$0.15   &  5.00$\pm$0.05   &  5.25$\pm$0.04                     & --               &5.33$\pm$0.04      & 5.46$\pm$0.04           & 5.50$\pm$0.30                                                          \\
Ar                      & 6.59$\pm$0.05   &  6.21$\pm$0.09   &  6.43$\pm$0.05                     & --               &6.50$\pm$0.05      & 6.62$\pm$0.05           & 6.40$\pm$0.13                                                   \\
Fe                      & 4.96$\pm$0.20   &  5.38$\pm$0.10   &  6.03$\pm$0.05                     & --               &5.86$\pm$0.10      & 5.99$\pm$0.10           & 7.50$\pm$0.04                                             \\
\noalign{\vskip3pt} \noalign{\hrule}\noalign{\vskip3pt}
\end{tabular}
\begin{description}
\item[$^a$] The He entries are before and after including corrections for
neutral helium respectively. The ICF scheme discussed in the text was used. For
the Core the abundances from FOS were used for carbon and oxygen and the mean
of VLT and FOS values were taken for nitrogen and neon; the other values were
adopted from the VLT data. For the Tail and Background the VLT values were
adopted; the carbon abundance for the Background was adopted from the M42
filament FOS position and may not be strictly representative of the local LV\,2
background. References: [1] This work; [2] Esteban et al. (2004) for two values
of the temperature fluctuation parameter, $t^2$ (e.g. Peimbert \& Costero
1969); [3] Asplund et al. (2009).
\end{description}
\end{table*}

In this Section elemental abundances in LV\,2 are derived using the direct
measurements for the (\elt, \eld) obtained above. In Table~5 ionic abundances
relative to \hp\ are listed for the Core, Tail and Background regions of LV\,2
based on CELs from the Argus data. The Core and Tail regions correspond to
background-subtracted spectra and so these measurements should reflect more
accurately the intrinsic composition of the proplyd.

Abundances were also computed from the Core, Tail and M42 filament FOS
observations. For the Core, background-subtracted line intensities were used in
order to be free from nebular contamination (see last column of Table~2). In
this case the Tail FOS spectrum was adopted as background due to the 1$''$
proximity of the respective FOS pointings. The two M42 filament observations
were not considered to be suitable for this purpose as they are 55~arcsec away
and the Orion Nebula surface brightness varies considerably over such scales.

As the precise density stratification of LV\,2 is unknown ionic abundances can
be sensitive to the choice of \eld\ for their respective emitting zones if one
relies only on lines of low \crd; lines of high \crd\ need to be considered for
consistency otherwise some ionic ratios could be underestimated (Rubin 1989).
In the determination of the \op/\hp, \opp/\hp, \np/\hp, and \sulp/\hp\ ratios
we also computed abundances from the $\lambda$2470, $\lambda$4363,
$\lambda$5755, and $\lambda$4069 lines respectively (all with \crd\ $>$ 10$^6$
\cmt). For instance, we found that regarding \opp, at 9000\,K and log[\eld
(\cmt)] $=$ 5.9, the difference in the ionic ratios from the $\lambda$5007 and
$\lambda$4363 lines is 10 per cent. For \np\ the difference between abundances
derived from the $\lambda$6584 and $\lambda$5755 lines is 30 per cent. For \op\
the $\lambda$3727 doublet returns a higher abundance ratio by 50 per cent
compared to the $\lambda$2470 line. In such cases the mean was taken. For
\sulp\ the value computed from the $\lambda$4069 line was adopted (from the VLT
data as the line is blended with \oii\ 3p--3d lines in the lower spectral
resolution FOS spectrum).

Abundances for \opp\ and \cpp\ were further computed from their respective
\oii\ and \cii\ RLs detected in the three co-added VLT spectra and are included
in Table~5. The following expression yields recombination line abundances
relative to \hp:

\begin{equation}\label{eq:icforlc}
\frac{~\rm {X^{i+1}}}{\rm H^+} = \frac{~\lambda}{4861.33}~\frac{\alpha_{\rm
eff} ({\rm H\beta})}{\alpha_{\rm eff}({\rm \lambda})}~\frac{I(\lambda)}{I({\rm
H\beta})},
\end{equation}\label{eq:icforlc}

\noindent where $I$($\lambda$) is the intensity at wavelength $\lambda$ (\AA)
of an RL of the recombining ion X$^{i+1}$, and $\alpha_{\rm eff}(\lambda)$,
$\alpha_{\rm eff} ({\rm H\beta})$ are the effective recombination coefficients
for the line in question and \hb, respectively. Recombination coefficients for
\oii\ and \cii\ were taken from Storey (1994) and Davey et al. (2000),
respectively. The \hep/\hp\ ratio was obtained from the \hei\ $\lambda$4471,
$\lambda$5876, $\lambda$6678 RLs (Table~6), using effective recombination
coefficients from Smits (1996) and correcting for the effects of collisional
excitation using the formulae in Benjamin, Skillman \& Smits (1999). A
correction for the presence of neutral helium was estimated following Peimbert,
Torres-Peimbert \& Ruiz (1992) so that:

\begin{equation}\label{eq:icforlc}
{\rm He/H} = {\rm He^{+}/H^{+}} \,\times\,\big(1 + \frac{{\rm S}^{+}}{{\rm S} -
{\rm S}^{+}}\big).
\end{equation}

An estimate of the abundances of elements heavier than He can be made via
ionization correction factor (ICF) methods based on standard schemes from the
literature. The results are listed in Table~7 for the Core, Tail and local
Background of LV\,2 (helium abundances with and without correction for He$^0$
are included). Corrections for the unobserved ions \npp, \nep, S$^{3+}$ and
Fe$^+$ were made as follows:

\begin{equation}\label{eq:icforlc}
{\rm N/H} = {\rm N^{+}/H^{+}} \,\times\,({\rm O}^{+} + {\rm O}^{2+})/{\rm
O}^{+},
\end{equation}

\begin{equation}\label{eq:icforlc}
{\rm Ne/H} = {\rm Ne^{2+}/H^{+}} \,\times\,({\rm O}^{+} + {\rm O}^{2+})/{\rm
O}^{2+},
\end{equation}

\begin{equation}\label{eq:icforlc}
{\rm S/H} = ({\rm S}^{+} + {\rm S}^{2+})/{\rm H}^{+} \,\times\,\big[1 -
(\frac{{\rm O}^{+}}{{\rm O}^{+} + {\rm O}^{2+}})^3\big]^{-1/3},
\end{equation}

\begin{equation}\label{eq:icforlc}
{\rm Fe/H} = 0.9\,\times\,({\rm O}^{+}/{\rm O}^{2+})^{0.08}\,\times\,({\rm
O}/{\rm O}^{+})\,\times\,{\rm Fe^{2+}/H^{+}},
\end{equation}

\noindent following Kingsburgh \& Barlow (1994), Peimbert \& Costero (1969),
Stasinska (1978), and Rodr\'{i}guez \& Rubin (2005), respectively. The
abundance of oxygen is the sum of \op/\hp\ and \opp/\hp\ ratios, and the
abundance of carbon is the sum of \cp/\hp\ and \cpp/\hp. For the determination
of oxygen and carbon recombination line abundances we relied on the \oii\ and
\cii\ RLs to obtain \opp/\hp\ and \cpp/\hp, but added in the \op/\hp\ and
\cp/\hp\ CEL values in each case (in the absence of \oi\ and C~{\sc i} RLs). No
UV carbon lines are detected by FOS from the Tail region (the \cii]
$\lambda$2326 detection is tentative) and so in estimating the carbon abundance
using solely the \cii\ $\lambda$4267 recombination line we corrected for the
presence of \cp\ adopting ICF(C) $=$ O/\opp\ (Kingsburgh \& Barlow 1994).
Finally, in the cases of chlorine and argon we corrected for the presence of
Cl$^+$ and Ar$^+$ adopting ICFs of 1.50$\pm$0.08 and 1.33$\pm$0.19,
respectively, from the Orion Nebula study of Esteban et al. (2004). In Fig.\,~8
the various abundance measurements for the species considered in this analysis
(except helium) are plotted.

\begin{figure*}
\centering \epsfig{file=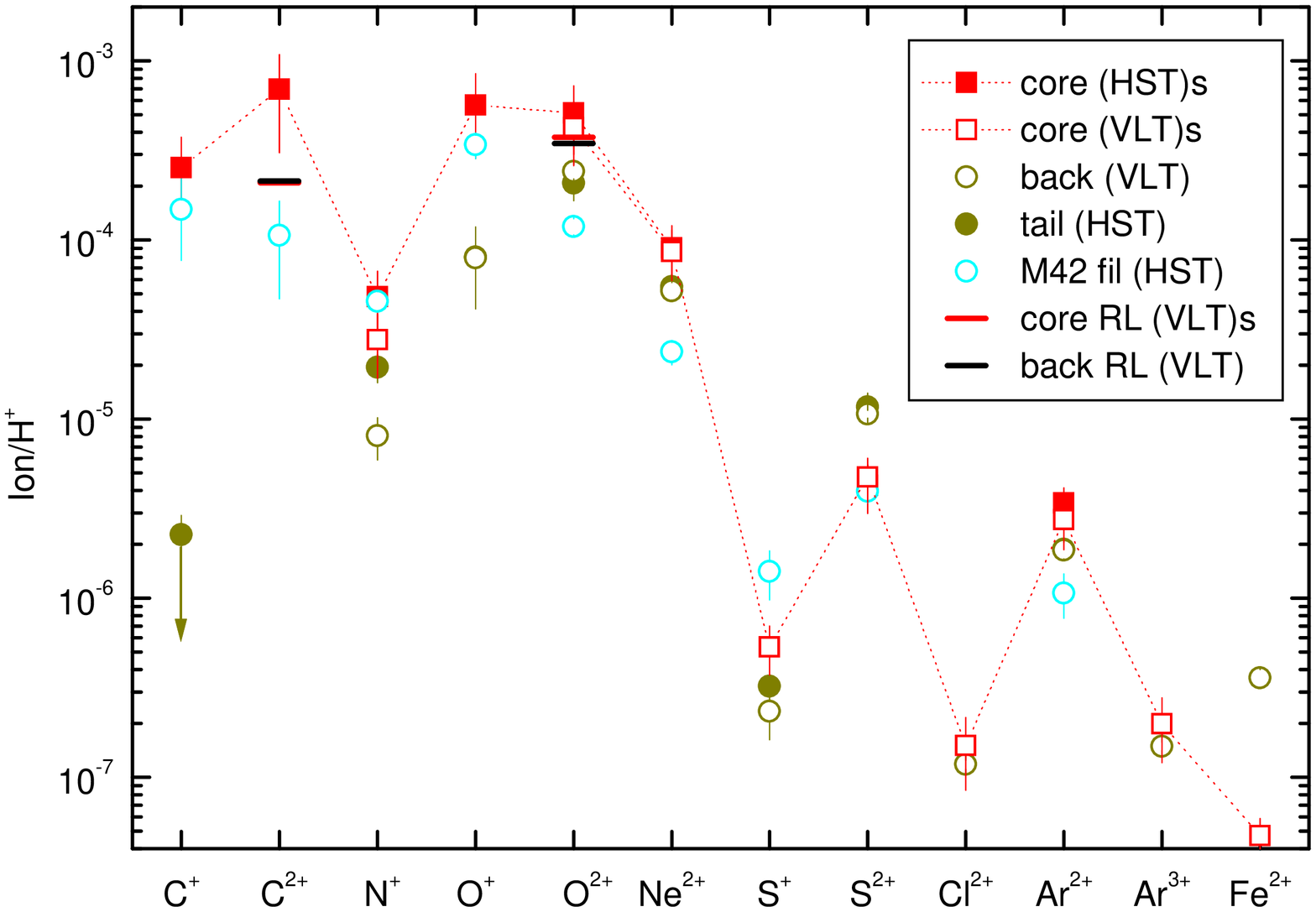, width=11 cm, scale=, clip=, angle=-90}
\caption{Ionic gas phase abundances in LV\,2 and in its Orion Nebula vicinity
from the present analysis. Abundances were computed from collisionally excited
lines. For \cpp\ and \opp\ abundances from recombination lines are shown as
horizontal bars. In the key box `s' refers to values from background subtracted
spectra.}
\end{figure*}

\subsection{Metallic recombination lines and the abundance anomaly}

\begin{figure}
\centering \epsfig{file=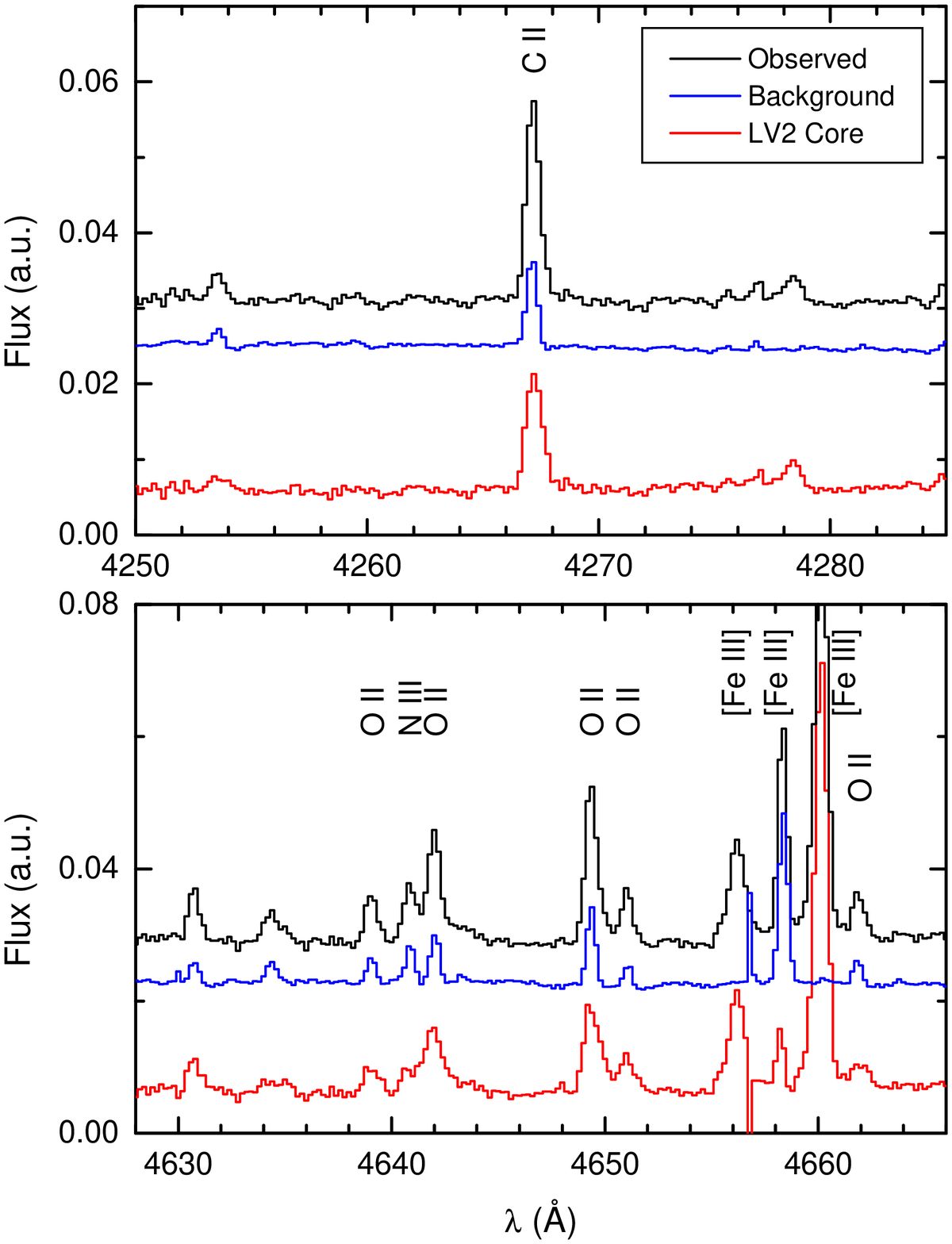, width=9 cm, scale=, clip=, angle=0}
\caption{The \cii\ 4267.15\,\AA\ 3d\,$^2$D -- 4f\,$^2$F$^{\rm o}$ and \oii\ V1
multiplet 3s\,$^4$P -- 3p\,$^4$D$^{\rm o}$ recombination lines on the VLT Argus
spectrum of LV\,2 at a resolution of 9.5 km s$^{-1}$ per pixel. Black: Observed
(proplyd $+$ M42 nebula), Blue: local M42 nebula background, and Red: LV\,2
Core (observed minus the background). Note the blue- and red-shifted components
of \ffeiii\ 4658\,\AA\ which arise in the outflow.}
\end{figure}

Recombination lines due to \oii\ 3s--3p and \cii\ 4f--3d transitions have been
recorded by VLT Argus arising from LV\,2 providing a measure of the \opp/\hp\
and \cpp/\hp\ abundance ratios independently of the \foiii\ and \fciii\ CELs.
The lines have previously been detected in Herbig-Haro objects in M42 (Blagrave
et al. 2006; Mesa-Delgado et al. 2009), as well as in the diffuse M42 gas (e.g.
Esteban et al. 2004). Here, several \oii\ V1 multiplet lines near 4650\,\AA\
are detected along with \cii\ $\lambda$4267.15 which is the strongest metallic
RL accessible in the optical (Fig.\,~9). Intensity maps of \oii\ $\lambda$4649
and \cii\ $\lambda$4267 are shown in Fig.\,~3. The lines are intrinsically
faint with intensities of less than one per cent of $I$(\hb) but offer the
advantage that the ratio of their emissivities to \hi\ lines is a weak function
of the plasma temperature and density (Storey 1994; Storey \& Hummer 1995; Liu
et al 1995; Davey et al. 2000; Tsamis et al. 2004). The derived abundances
should thus be less prone to errors resulting from uncertainties in the
measurement of these quantities (e.g. Peimbert et al. 1993; Tsamis et al 2003a;
Esteban et al 2004). This is in contrast to abundances derived from ratios of
CELs to \hi\ lines which have an exponential sensitivity to the electron
temperature, and also depend on \eld\ when CELs of low \crd\ are used (e.g.
Rubin 1989; Tsamis et al. 2003b).

The well-documented abundance anomaly in \hii\ regions refers to the fact that
higher abundances of \opp/\hp\ and \cpp/\hp\ are derived from \oii\ and \cii\
RLs than from the \foiii\ and \fciii\ CELs of the same ions (e.g. Peimbert et
al. 1993; Tsamis et al. 2003a; Mesa-Delgado \& Esteban 2010). This has
repercussions for the representative abundances of O/H and C/H in the nebulae
and by extension for the metallicity of the galaxies that host them. The ratio
of RL to CEL abundance determinations for a given ion (the so-called abundance
discrepancy factor, ADF) takes values from $\sim$2 to 5 in \hii\ regions, and
is best measured for \opp\ which has RLs (e.g. 4649.13\,\AA) and CELs (e.g.
4958.91, 5006.84\,\AA) falling in the blue part of the spectrum that can be
observed simultaneously with the same telescope/spectrograph configuration. It
is more difficult to obtain the ADF for \cpp\ which has a CEL in the UV (the
\fciii\ 1908\,\AA\ intercombination doublet) and a RL in the blue (\cii\
4267\,\AA) and be certain that the same volume of nebula is sampled as data
obtained from different telescopes/apertures need to be combined.

We take advantage of the co-spatial coverage over LV\,2 of \fciii\
$\lambda$1908 and \cii\ $\lambda$4267 with FOS and Argus respectively to
measure both the \cpp\ and \opp\ ADFs with a good degree of confidence. In
Table~8 the values that we obtained are listed for the LV\,2 Core, Tail and Tip
positions and for the local nebula background in the form of

\begin{equation}\label{eq:icforlc}
{\rm ADF(X^{2+})} \equiv \frac{({\rm X^{2+}/H^{+}})_{\rm RL}}{({\rm
X^{2+}/H^{+}})_{\rm CEL}}.
\end{equation}

\noindent The quoted uncertainties are not due to formal errors but are
associated with the variation of the quantity arising from the range of the
CEL-computed abundances (with their known sensitivity to \elt\ and \eld).
Values are tabulated for the background-subtracted spectra but also for the
observed spectra which include the M42 nebular emission, so as to examine the
difference in the numbers obtained. The \opp/\hp\ and \cpp/\hp\ RL ratios in
Table~5 were derived from the VLT data. Additionally, the 12$+$log(\opp/\hp)
and 12$+$log(\cpp/\hp) RL ratios corresponding to the \emph{observed} Core VLT
spectrum are 8.65$\pm$0.05, 8.37$\pm$0.04, while those from the \emph{observed}
Tail VLT spectrum are 8.57$\pm$0.05 and 8.35$\pm$0.04 respectively. These along
with the corresponding CEL values listed in the footnotes of Table~8 were used
to obtain the ADFs corresponding to the intrinsic as well as the
nebula-contaminated (observed) LV\,2 spectra.

\setcounter{table}{7}
\begin{table}
\caption{Abundance discrepancy factors: the ratio of RL to CEL abundance
determinations for \opp\ and \cpp.$^a$}
\begin{tabular}{lcc}
\noalign{\vskip3pt} \noalign{\hrule} \noalign{\vskip3pt}
                    &ADF(\opp)                   &ADF(\cpp)                     \\
&\multicolumn{2}{c}{Background-subtracted spectra}\\
Core (VLT)          & 0.8$^{+0.6}_{-0.3}$     & --                            \\
Core (FOS)          & 0.7$^{+0.5}_{-0.3}$     & 0.3$^{+0.2}_{-0.1}$       \\
Tail (VLT)          & 2.6$^{+1.1}_{-0.8}$     & --                          \\
Tip (FOS)           & 0.7$^{+0.4}_{-0.3}$     & 0.3$^{+0.3}_{-0.1}$       \\ 
&\multicolumn{2}{c}{Observed spectra}\\
Core  (VLT)         & 1.9$^{+3.1}_{-0.7}$     & --                            \\ 
Core (FOS)          & 2.1$^{+2.5}_{-0.8}$     & 1.2$^{+3.2}_{-0.5}$      \\   
Tail (FOS)          & 1.8$^{+0.3}_{-0.3}$     & --                           \\
Tip (FOS)           & 3.3$^{+1.0}_{-0.6}$     & 3.9$^{+2.2}_{-1.0}$     \\ 
\noalign{\vskip2pt}
Background (VLT)    & 1.5$^{+0.6}_{-0.4}$     &  --                          \\

\noalign{\vskip3pt} \noalign{\hrule}\noalign{\vskip3pt}
\end{tabular}
\begin{description}
\item[$^a$] The \cpp/\hp\ and \opp/\hp\ RL ratios were obtained from the VLT
spectra in all cases (Table~3). In units where log\,H $=$ 12 the corresponding
CEL abundance ratios used to compute the ADFs are: for the
background-subtracted spectra of Core (VLT), Core (FOS), Tail (VLT) see
Table~5; for Tip (FOS): \opp\ $=$ 8.75, \cpp\ $=$ 8.85 (at \elt\ $=$ 9000\,K,
log[\eld\ (\cmt)] $=$ 5.81). For the observed spectra -- Core (VLT): \opp\ $=$
8.38$^{+0.21}_{-0.43}$ (at \elt\ $=$ 9650$\pm$1350\,K, log[\eld\ (\cmt)] $=$
5.29$^{+0.29}_{-1.48}$) ; Core (FOS): \opp\ $=$ 8.33$^{+0.19}_{-0.61}$, \cpp\
$=$ 8.29$^{+0.24}_{-0.57}$ using data from Table~2; Tail (FOS) see Table~5; Tip
(FOS): \opp\ $=$ 8.13$^{+0.08}_{-0.11}$, \cpp\ $=$ 7.79$^{+0.13}_{-0.20}$ using
data from Table~2; Background (VLT) see Table~5.

\end{description}
\end{table}

\section{Discussion and conclusions}

The gas-phase abundances of light metals in LV\,2 measured via UV and optical
emission line spectroscopy are generally higher than those in the Orion Nebula.
The differences are the largest for carbon and oxygen when comparing values
obtained from CELs. The differences become smaller when the recombination line
results are compared: $\Delta$C $=$ $+$0.11 and $\Delta$O $=$ $+$0.31 dex.
Assuming that the RL values for these two elements are more reliable then LV\,2
has a carbon abundance a factor of 1.3--1.7 higher than M42 (when comparing
with the local background or with Esteban et al. 2004), and a factor of 1.7
higher than the solar value by Asplund et al. (2009). If some of the C is
locked in grains, particularly in the inner disk regions, then the difference
with the sun would be even higher. For nitrogen there are no substantial
differences between LV\,2, M42 (for $t^2$ $=$ 0.022) or the sun. The oxygen
abundance in LV\,2 is higher by a factor of $\sim$2 compared to the local
background, M42, or the sun. Again the difference with the sun would be greater
if the unknown amount of oxygen incorporated in solids was considered. Neon,
which as a noble gas is not expected to be affected by depletion on grains, is
also overabundant by a factor of 2.2--2.6 compared to the sun and the local
nebula, or a factor or 1.9 when comparing with the recent determination in M42
of 8.01$\pm$0.01, based on {\it Spitzer} data (Rubin et al. 2010). The solar Ne
abundance is however controversial as Drake \& Testa (2005) have advocated a
value $\sim$2.5 times higher than Asplund et al. (2009); this would bring the
LV\,2 and solar neon abundances into agreement.

The oxygen and neon abundances in LV\,2 are consistent with the range of
abundances in the X-ray bright coronae of 35 pre-main sequence Orion Nebula
Cluster stars observed by Maggio et al. (2007); the oxygen abundance is
$\approx$0.2 dex higher than in 13 early B-type stars of the Ori OB1
association studied by Simon-Diaz (2010).

The sulphur abundance is consistent with the M42 value within the
uncertainties. Chlorine too is consistent with the M42 value, and the sun. The
argon abundance is in agreement with that in B-type stars of the Orion
association (6.66$\pm$0.06; Lanz et al. 2008), and in M42. The iron abundance
is most certainly severely underestimated as the important Fe$^+$ ion is not
observed, and a large percentage of the total must be in the dust phase; the
same is true for the local nebula as the expected dominant Fe$^{3+}$ ion (e.g.
Rubin et al. 1991) is not observed here. As a result, the iron abundance
estimates for LV\,2 and M42 cannot be directly compared with the sun.

The second noteworthy result from this study is that the abundance anomaly,
classically manifested by ADF(X$^{+i}$) $>$ 1.0, goes away when one considers
the intrinsic spectrum of the proplyd, that is, with the foreground/background
nebula contamination removed (Table~8)\footnote{With the exception of the Tail
region where the background subtraction was probably not as accurate due to its
extended nature.}. The upward correction overwhelmingly affects the temperature
and density dependent CEL diagnostics, not the RLs. In this case the RL and CEL
abundances for oxygen come into very good agreement for the Core and Tip
regions of LV\,2, which encompass a large part of the luminous flux from the
object. The carbon abundance derived from the UV CELs is then a factor of 2--3
higher that the RL value. This however could be due to an underestimation of
the \elt\ in the zone emitting the 1908\,\AA\ line, the most temperature
sensitive line in this study; adopting a value in the upper \elt\ range
($\sim$9700\,K) the CEL and RL measurements would come into agreement.

These effects on the ADF are due to the disproportionate bias that the \hii\
region contamination imparts on the CEL temperature and abundance diagnostics
compared to the RL diagnostics: the $I_{\rm sum V1}$(\oii)/$I$(\hb) and
$I$(\cii\ $\lambda$4267)/$I$(\hb) intensity ratios are practically the same in
the intrinsic LV\,2 core spectrum and in the LV\,2 core $+$ M42 spectrum. In
this sense the proplyd is not a prolific metallic RL emitter such as the inner
regions of planetary nebulae (e.g. Tsamis et al. 2008). However considering the
high emission measure ($\int$\eld\ $N_{\rm H}~dV$) of the skin-deep but very
dense photoionized layer of the proplyd one could instead say that LV\,2 is at
least as prolific a \cii, \oii\ RL emitter as the diffuse Orion Nebula along
the same line of sight. At the same time, the proplyd appears hotter, less
dense and with a lower (forbidden-line) oxygen and carbon abundance when the
M42 contribution to the relevant CELs is neglected than when this contribution
is subtracted.

If one considers that the decades old ADF problem has been solved in the case
of LV\,2 `simply' by subtracting the \hii\ region spectrum and examining in
detail the spectrum of one dense clump where the margin of uncertainty in the
derived quantities is reduced, then one is impelled to acknowledge the
following:

(i) In the presence of a population of small and dense partially-ionized clumps
along a random line of sight, the CEL-derived Orion Nebula gaseous abundances
would be lower limits based on the principles discussed above (cf. also Rubin
1989; Viegas \& Clegg 1994). In the vicinity of LV\,2 the gas-phase M42
abundances would in fact lie somewhere between the LV\,2 Core and local
background values.

(ii) The simplest solution to the so-called ADF/$t^2$ problem encountered in
\hii\ regions would then be one where density inhomogeneities are playing havoc
with the classic forbidden line diagnostics.

(iii) The metallicity of nebulae where spatially resolved observations are
impossible, such as extragalactic \hii\ regions, will need to be corrected
upwards. Whether there will be repercussions for cosmic chemical evolution
studies will depend on the correction factor and how that factor varies with
metallicity or other properties of the ionized interstellar medium, such as the
degree of `clumpiness'.

\section*{Acknowledgments}

We wish to thank the FLAMES support astronomers at ESO for scheduling the VLT
service mode observations. Thanks also to Bob Rubin, Bob O'Dell, and the
referee for helpful comments. This research has used \hst\ data obtained from
the Space Telescope European Coordinating Facility's archive at ESO. YGT and
JMV acknowledge funding from grants AYA2007-67965-C03-02/CSD2006-00070
(Ingenio-Consolider 2010) of the Spanish Ministry of Science and Innovation.

YGT further acknowledges the award of a Marie Curie intra-European Fellowship
within the 7$^{\rm th}$ European Community Framework Programme (grant agreement
PIEF-GA-2009-236486).

\label{lastpage}

\end{document}